\documentclass[useAMS,usenatbib,fleqn]{mnras}
\usepackage{amsmath}
\usepackage{graphicx}
\usepackage{color}
\usepackage{natbib}
\usepackage{amssymb}

\bibpunct{(}{)}{;}{a}{}{,}

\def\eps{{\epsilon}}

\title{Intrinsic alignment of redMaPPer clusters: cluster shape - matter density correlation}

\author[Edo van Uitert and Benjamin Joachimi]
{Edo van Uitert$^{1}$\thanks{vuitert@ucl.ac.uk}, Benjamin Joachimi$^1$ \\ 
$^1$ Department of Physics and Astronomy, University College London, Gower Street, London WC1E 6BT, UK \\ 
}

\pubyear{2017}

\begin{document}

\maketitle

\begin{abstract}
We measure the alignment of the shapes of galaxy clusters, as traced by their satellite distributions, with the matter density field using the public redMaPPer catalogue based on SDSS-DR8, which contains $26\,111$ clusters up to $z\sim0.6$. The clusters are split into nine redshift and richness samples; in each of them we detect a positive alignment, showing that clusters point towards density peaks. We interpret the measurements within the tidal alignment paradigm, allowing for a richness and redshift dependence. The intrinsic alignment (IA) amplitude at the pivot redshift \mbox{$z=0.3$} and pivot richness $\lambda=30$ is $A_{\rm IA}^{\rm gen}=12.6_{-1.2}^{+1.5}$. We obtain tentative evidence that the signal increases towards higher richness and lower redshift. Our measurements agree well with results of maxBCG clusters and with dark-matter-only simulations. Comparing our results to IA measurements of luminous red galaxies, we find that the IA amplitude of galaxy clusters forms a smooth extension towards higher mass. This suggests that these systems share a common alignment mechanism, which can be exploited to improve our physical understanding of IA.
\end{abstract}

\begin{keywords}
large-scale structure of Universe - galaxies: clusters: general - dark matter - methods: statistical - methods: data analysis
\end{keywords}

\maketitle


\section{Introduction}
Galaxies inside dark matter haloes are subject to gravitational tidal fields from the large-scale distribution of matter. Additionally, new material is continuously accreted onto the haloes along preferred directions. The consequence is that galaxy shapes become aligned with the density field. Neighbouring galaxies are aligned along similar directions and hence their observed shapes are correlated, which is known as intrinsic alignment \citep[IA; for recent reviews, see][]{Troxel15,Joachimi15,Kiessling15,Kirk15}. The IA signal contains information about galaxy formation processes, but is mainly studied for a different reason: it is a major contaminant of the cosmic shear signal in future lensing surveys such as Euclid \citep{Laureijs11}, LSST \citep{LSST09} and WFIRST \citep{Spergel15}. If unaccounted for, IA will significantly bias cosmological inferences \citep{Kirk10,Kirk12,Kirk15,Krause16}. Hence, accurate and precise IA models are needed. \\
\indent The IA signal has been measured in observations \citep[e.g.][]{Mandelbaum06,Okumura09,Hirata07,Joachimi11,Li13,Singh15}, in $N$-body simulations \citep[e.g.][]{Croft00,Hopkins05} and in hydrodynamical simulations \citep{Codis15,Velliscig15,Chisari15,Chisari16,Tenneti16,Hilbert16}. The picture that is emerging is that massive, red galaxies are pointing towards matter overdensities, which can be reasonably well described by the tidal alignment model \citep{Catelan01,Hirata04,Bridle07,Blazek11,Blazek15} at large scales. For blue galaxies, IA has not yet been clearly detected \citep{Hirata07,Mandelbaum11}. \\
\indent To gain further observational input for IA models and extend the mass range, we investigate the alignment between the shapes of galaxy clusters and the density field. If clusters are subject to the same physical mechanisms that intrinsically align galaxies, they could be used to improve IA models of galaxies, to the benefit of the exploitation of future cosmic shear surveys. \\
\indent A positive shape - density correlation has been observed in the Sloan Digital Sky Survey (SDSS) for galaxy groups \citep{Wang09,Paz11}. For galaxy clusters, \citet{Smargon12} obtained a clear detection using two cluster samples up to transverse separations of 100 Mpc/$h$. The amplitude of the correlation was reported to be significantly lower than predictions from numerical simulations based on a $\Lambda$CDM cosmology \citep{Hopkins05}, and \citet{Smargon12} argued that various systematic observational uncertainties may lie at the root of this. \\
\indent In this work, we use the publicly available redMaPPer cluster catalogue \citep{Rykoff14}, which contains over three times more clusters than the maxBCG sample used in \citet{Smargon12}, and extends to higher redshifts. This enables us to study the cluster shape - density correlation as a function of the number of cluster members (i.e. the richness) and redshift. Furthermore, it provides us with a new sample to revisit the previously reported tension with numerical simulations. \\
\indent The outline is as follows. We describe the theoretical background and the estimators that we use in Sect. \ref{sec_meth}. The main aspects of the redMaPPer cluster sample are briefly discussed in Sect. \ref{sec_data}. We present our measurements and discuss the results in Sect. \ref{sec_res}, and we conclude in Sect. \ref{sec_conc}. Throughout the paper we assume a standard $\Lambda$CDM cosmology with $\Omega_{\Lambda}=0.73$, $\Omega_{\rm M}=0.27$, $\sigma_8 =0.8$, $n_s=1.0$ and $h=0.7$ the dimensionless Hubble parameter, which is consistent with the best-fitting cosmological parameters from WMAP9 \citep{Hinshaw13}. We adopt these slightly outdated parameters as it eases a comparison with previous papers. All distances are in co-moving (not physical) units and are presented in units of Mpc/$h$.


\section{Methodology}\label{sec_meth}
To measure the correlation between the shapes of galaxy clusters and the density field (traced by the distribution of the same clusters), we adopt the following estimator:
\begin{equation}
\widehat{\xi}_{\rm g+}(R_{\rm p},\Pi)=\frac{S_+ D_{\rm d}}{D_{\rm s} D_{\rm d}} - \frac{S_+ R}{D_{\rm s} R} \;,
\end{equation}
with $S_+ D_{\rm d}$ the correlation between cluster shapes and the density sample, $D_{\rm s} D_{\rm d}$ the number of cluster shape - density pairs, $S_+ R$ the correlation between cluster shapes and random points and $D_{\rm s} R$ the number of cluster shape - random point pairs. $R_{\rm p}$ and $\Pi$ are the comoving transverse and line-of-sight distances, respectively. The $S_+ R$ term removes potential systematics in the cluster shape sample. We have checked that this signal is consistent with zero, but still subtract it as it decreases the measurement error \citep[see also][]{Singh16}. We note that this estimator is identical to the ones used in \citet{Joachimi11} and \citet{Singh15}, although we write it differently to explicitly account for the different number of clusters in a shape sample and the number of random points, which is implicitly assumed in their equations. The total projected IA signal is obtained by integrating along the line of sight:
\begin{equation}
\widehat{w}_{\rm g+}(R_{\rm p}) = \int_{\Pi_{\rm min}}^{\Pi_{\rm max}} {\rm d}\Pi \; \widehat{\xi}_{\rm g+}(R_{\rm p},\Pi) \;.
\label{eq_wg+m}
\end{equation}
In practise, this integral turns into a sum over correlation functions binned in ranges of $\Pi$. Ideally, one would like to adopt $\Pi_{\rm min}=-\infty$ and $\Pi_{\rm max}=\infty$, but at large line-of-sight separations $\widehat{\xi}_{g+}(R_{\rm p},\Pi)$ is so small that effectively only noise is added, which makes the IA signal increasingly noisy. On the other hand, redshift space distortions and photometric redshift errors spread the signal in the radial direction, and the measured signal will be biased if the integration ranges are not large enough to enclose this smearing. Therefore, it is optimal to choose $\Pi_{\rm min}$ and $\Pi_{\rm max}$ such that they cover a range that is as small as possible, under the condition that the induced bias due to any missed signal is much smaller than the statistical errors. We detail our choices in Sect. \ref{sec_data}.\\
\indent We also measure the clustering signal of the redMaPPer clusters using the LS \citep{Landy93} estimator, which we need to constrain the cluster bias:
\begin{equation}
\widehat{\xi}_{\rm gg}(R_{\rm p},\Pi)=\frac{D_{\rm d} D_{\rm d} - 2 D_{\rm d} R + R R}{R R} \;,
\end{equation}
where $D_{\rm d} D_{\rm d}$ indicates the number of cluster pairs (our density sample), $D_{\rm d} R$ the number of cluster - random point pairs, and $R R$ the number of random point pairs. The counts with random points are scaled with the ratio of the total number of clusters and the total number of random points. Each random point has a weight to account for its detection probability \citep{Rykoff16}, which we include in the pair counts. We obtain the total projected clustering signal by integrating along the line of sight,
\begin{equation}
\widehat{w}_{\rm gg}(R_{\rm p}) = \int_{\Pi_{\rm min}}^{\Pi_{\rm max}} {\rm d}\Pi \; \widehat{\xi}_{\rm gg}(R_{\rm p},\Pi) \;,
\label{eq_wggm}
\end{equation}
using the same integral limits as in Eq. (\ref{eq_wg+m}). \\
\indent We interpret the intrinsic alignment signal with the following model \citep{Hirata04,Joachimi11}:
\begin{equation}
w_{\rm g+}(R_{\rm p}) = -b_{\rm g} \int {\rm d}z \; W(z) \int_0^\infty \frac{{\rm d} k_\perp \; k_\perp}{2\pi} J_2 (k_\perp R_{\rm p}) P_{\delta I}(k_\perp,z)  \;,
\end{equation}
with $b_{\rm g}$ the cluster bias, which we assume to be scale-independent, $k_\perp$ the wavenumber transverse to the line of sight, $J_2$ the second Bessel function of the first kind, $P_{\delta I}$ the IA power spectrum, and 
\begin{equation}
W(z)=\frac{p^2(z)}{\chi^2(z) \chi'(z)} \left[ \int {\rm d}z \; \frac{p^2(z)}{\chi^2(z) \chi'(z)} \right]^{-1} \;.
\end{equation}
Here, $p(z)$ indicates the redshift probability distribution of the redMaPPer cluster sample, $\chi(z)$ denotes the comoving distance and $\chi'(z)$ the derivative of the comoving distance with respect to redshift. $W(z)$ accounts for the fact that the number of pairs is proportional to the comoving volume, while the integral is performed as a function of redshift \citep{Mandelbaum11}.\\
\indent We model $P_{\delta I}$ using the linear alignment model \citep{Catelan01,Hirata04,Hirata10}:
\begin{equation}
P_{\delta I}(k,z) = -A_{\rm IA} C_1 \rho_{\rm crit} \frac{\Omega_{\rm M}}{D(z)} P_\delta(k,z) \;,
\label{eq_pi}
\end{equation}
with $D(z)$ the growth factor, normalised to unity at $z=0$, $\rho_{\rm crit}$ the critical density and $P_\delta(k,z)$ the matter power spectrum. The power spectrum is computed for our fiducial cosmological parameters, using the transfer function of \citet{Eisenstein98} and including the non-linear correction of \citet{Smith03}; again, we adopt these slightly outdated prescriptions to ease comparison with results from the literature. The normalization is absorbed into $C_1$, such that $C_1 \rho_{\rm crit} \approx 0.0134$. To account for a possible redshift and richness dependence of the IA signal, we generalize Eq. (\ref{eq_pi}) to:
\begin{equation}
P_{\delta I}(k,z) = -A_{\rm IA}^{\rm gen} C_1 \rho_{\rm crit} \frac{\Omega_{\rm M}}{D(z)} P_\delta(k,z) \left ( \frac{1+z}{1+z_0} \right )^\eta \left ( \frac{\lambda}{\lambda{_0}} \right )^\beta \;,
\label{eq_pi2}
\end{equation}
with $\lambda$ the mean richness of the sample, $z_0$ a pivot redshift fixed to 0.3 and $\lambda_0$ a pivot richness fixed to 30. \\
\indent The amplitude $A_{\rm IA}$ and the bias $b_{\rm g}$ are completely degenerate. To lift the degeneracy, we fit the clustering signal of the cluster sample, which provides independent constraints on $b_{\rm g}$, using the following model that accounts for redshift space distortions \citep{Joachimi15b}:
\begin{eqnarray}
w_{\rm gg}(R_{\rm p}) = 2 \; b_{\rm g}^2 \int_0^\infty {\rm d}z \; W(z) \sum_{l=0}^{2} \alpha_{2l} \left[\frac{f(z)}{b_{\rm g}} \right] \int_0^{\Pi_{\rm max}} {\rm d}\chi \nonumber\\ 
\times \; \xi_{\delta\delta,2l}\left( \sqrt{\chi^2+R_{\rm p}^2},z \right) L_{2l} \left(\frac{\chi}{\sqrt{\chi^2+R_{\rm p}^2}}\right) + C_{\rm IC} \;,
\label{eq_wgg}
\end{eqnarray}
with $f(z)$ the growth rate, $\alpha_{2l}(f(z)/b_{\rm g})$ polynomials which are given in Eq. (48) to (50) of \citet{Baldauf10}, $\xi_{\delta\delta,2l}(R,z)$ are the multipoles of the matter correlation function, which are computed as
\begin{equation}
\xi_{\delta\delta,2l}(R,z)=\frac{(-1)^l}{2\pi^2}\int_0^\infty {\rm d}k \; k^2 j_{2l}(kR) P_\delta(k,z) \;,
\end{equation}
with $j_{2l}$ the spherical Bessel functions. Furthermore, $L_{2l}$ indicate the Legendre polynomials and $C_{\rm IC}$ the integral constraint \citep{Roche99}, which  accounts for the bias in the observed clustering signal that is caused by the use of a finite survey area. We estimated $C_{\rm IC}$ using the random pair counts and found $C_{\rm IC}<1$, much smaller than the observed signal, hence we did not include this term in the modeling. Since the clustering signal is fit to scales that include the quasi-linear regime, $b_{\rm g}$ should be regarded as an effective linear bias. \\
\indent In summary, the general IA model has three free parameters: $A_{\rm IA}^{\rm gen}$, $\eta$ and $\beta$.  We assume flat, uninformative priors in the fit with ranges $A_{\rm IA}^{\rm gen}\in[0;30]$, $\eta \in[-10;10]$ and $\beta \in[-5;5]$. We also perform a run in which we fit the $A_{\rm IA}$ of each subsample separately, where we adopt a flat prior in the range $A_{\rm IA} \in[0;100]$. The bias of the density sample, $b_{\rm g}$, once obtained by fitting the clustering signal, is held fixed to its best-fitting value when we fit the IA signals. Not propagating the errors on $b_{\rm g}$ is safe, because they are much smaller than the errors on the other fit parameters.


\section{Data}\label{sec_data}
To measure the cluster shapes and to define the density field, we used the redMaPPer cluster catalogue \citep{Rykoff14} version 6.3, which has been made publicly available\footnote{http://risa.stanford.edu/redmapper/}. In short, the redMaPPer cluster finder uses photometric data from SDSS-DR8 \citep{Aihara11} to find clusters with the red-sequence technique using an iterative approach. The algorithm first calibrates the model red sequence as a function of redshift. This calibrated model is used to identify spatial overdensities of red-sequence candidates, which in turn is used to calibrate the model red sequence again. A detailed description of the cluster finder algorithm and its performance can be found in \citet{Rykoff14,Rozo14,Rozo15a,Rozo15b} \\
\indent The cluster finder assigns probabilities to the top five potential brightest cluster galaxies (BCGs) of being the cluster centre. Furthermore, each cluster member candidate is assigned a probability $p_{\rm mem}$ of belonging to the cluster. The cluster redshift is estimated by combining the red-sequence redshift estimates of cluster members with $p_{\rm mem}>0.9$. Finally, the cluster richness $\lambda$ is estimated as the sum of $p_{\rm mem}$ of all candidate members. \\
\begin{figure}
   \centering
   \includegraphics[width=0.95\linewidth,angle=270]{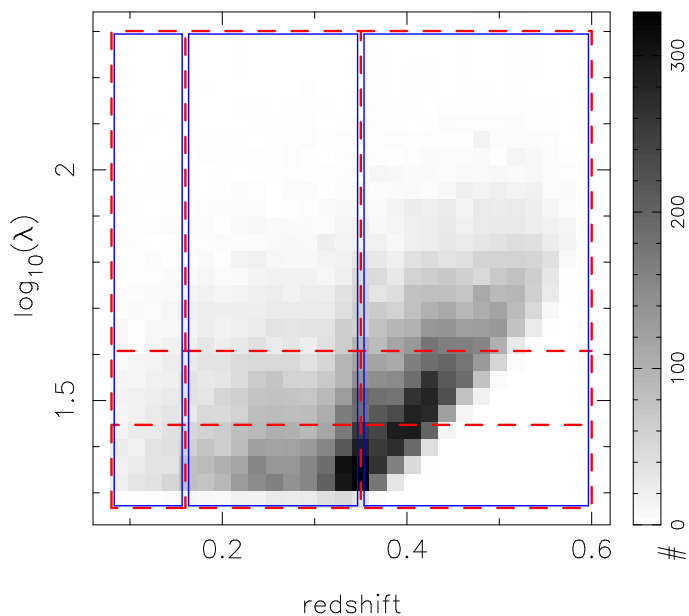}
   \caption{Distribution of cluster richnesses and redshifts for the redMaPPer cluster sample. The red dashed boxes indicate the selection of the nine shape samples, while the blue solid boxes illustrate the selection of the three density samples.}
   \label{plot_lamz}
\end{figure}
\begin{table*}
  \centering
  \caption{Properties of the cluster sample. The first column indicates the redshift cut, the second column the richness cut, the third column the number of clusters, the fourth column the mean redshift, the fifth column the mean richness, the sixth column the mean cluster ellipticity, the seventh column the cluster bias (determined using all redMaPPer clusters at that redshift) and the eighth column the amplitude of the linear alignment model.}
  \begin{tabular}{c c c c c c c c} 
  \hline
Redshift cut & Richness cut &  ${\rm N_{clus}}$ & $\langle z \rangle$ &  $\langle \lambda \rangle$ & $\langle \epsilon_{\rm clus} \rangle$ & $b_{\rm g}$ & $A_{\rm IA}$ \\
  \hline\hline
\\
$0.08<z\leq0.16$ & $19.8<\lambda\leq28$     & 490 & 0.127 & 23.7 & 0.160 &  & 16.2 $\pm$ 11.8 \\
$0.08<z\leq0.16$ & $28<\lambda\leq40.5$ & 301 & 0.127 & 33.2 & 0.164 & 4.22 $\pm$ 0.35 & 48.0 $\pm$ 22.0 \\
$0.08<z\leq0.16$ & $\lambda>40.5$      & 206 & 0.127 & 58.2 & 0.134 & & 36.9 $\pm$ 11.2 \\
\\
$0.16<z\leq0.35$ & $19.8<\lambda\leq28$     & 4634 & 0.275 & 23.4 & 0.129 & & 10.4 $\pm$ 2.6 \\
$0.16<z\leq0.35$ & $28<\lambda\leq40.5$ & 2628 & 0.273 & 33.2 & 0.122 & $4.25_{-0.16}^{+0.15}$ & 15.6 $\pm$ 3.0\\
$0.16<z\leq0.35$ & $\lambda>40.5$      & 1609 & 0.272 & 57.9 & 0.112 & & 19.1 $\pm$ 3.2 \\
\\
$0.35<z\leq0.60$ & $19.8<\lambda\leq28$     & 3077 & 0.383 & 24.5 & 0.122 & & 11.0 $\pm$ 3.1 \\
$0.35<z\leq0.60$ & $28<\lambda\leq40.5$ & 5371 & 0.420 & 33.8 & 0.116 & 4.61 $\pm$ 0.27 & 10.9 $\pm$ 2.4\\
$0.35<z\leq0.60$ & $\lambda>40.5$      & 6460 & 0.465 & 59.1 & 0.112 & & 15.1 $\pm$ 2.4 \\
  \hline
  \end{tabular}
  \label{tab_clusprop}
\end{table*} 
\indent The shapes of the clusters are determined using the distribution of satellites. Satellites are expected to trace the overall dark matter distribution \citep{Kang07,Agustsson10,Dong14,Wang14} and its orientation \citep{Evans09,VanUitert16}. Hence the shape of clusters as traced by their satellite distribution is expected to exhibit a similar IA effect as the dark matter. We determined the projected moments of the cluster member distribution using all cluster member candidates with $p_{\rm mem}>0.2$, and weighing the selected cluster members with their $p_{\rm mem}$:
\begin{equation}
Q_{ij} =\frac{\sum_k (\theta_{i,k}-\theta_{i}^{\rm BCG})(\theta_{j,k}-\theta_{j}^{\rm BCG}) p_{{\rm mem},k}}{\sum_k p_{{\rm mem},k}}, \hspace{2mm} i,j \in \{1,2\},
\label{eq_mom}
\end{equation}
where the sum runs over all cluster members, $(\theta_{1,k},\theta_{2,k})$ is the angular position of cluster member $k$  and $(\theta_1^{\rm BCG},\theta_2^{\rm BCG})$ is the position of the most likely BCG. From these moments we formed the complex ellipticity of each cluster:
\begin{equation}
\epsilon_{\rm clus}=\frac{Q_{11}-Q_{22}+2{\rm i}Q_{12}}{Q_{11}+Q_{22}+2\sqrt{Q_{11}Q_{22}-Q_{12}^2}} \;. \\
\label{eq_eclus}
\end{equation}
This procedure is similar to the one outlined in \citet{Huang16}. \\
\begin{figure*}
   \centering
   \includegraphics[width=1\linewidth]{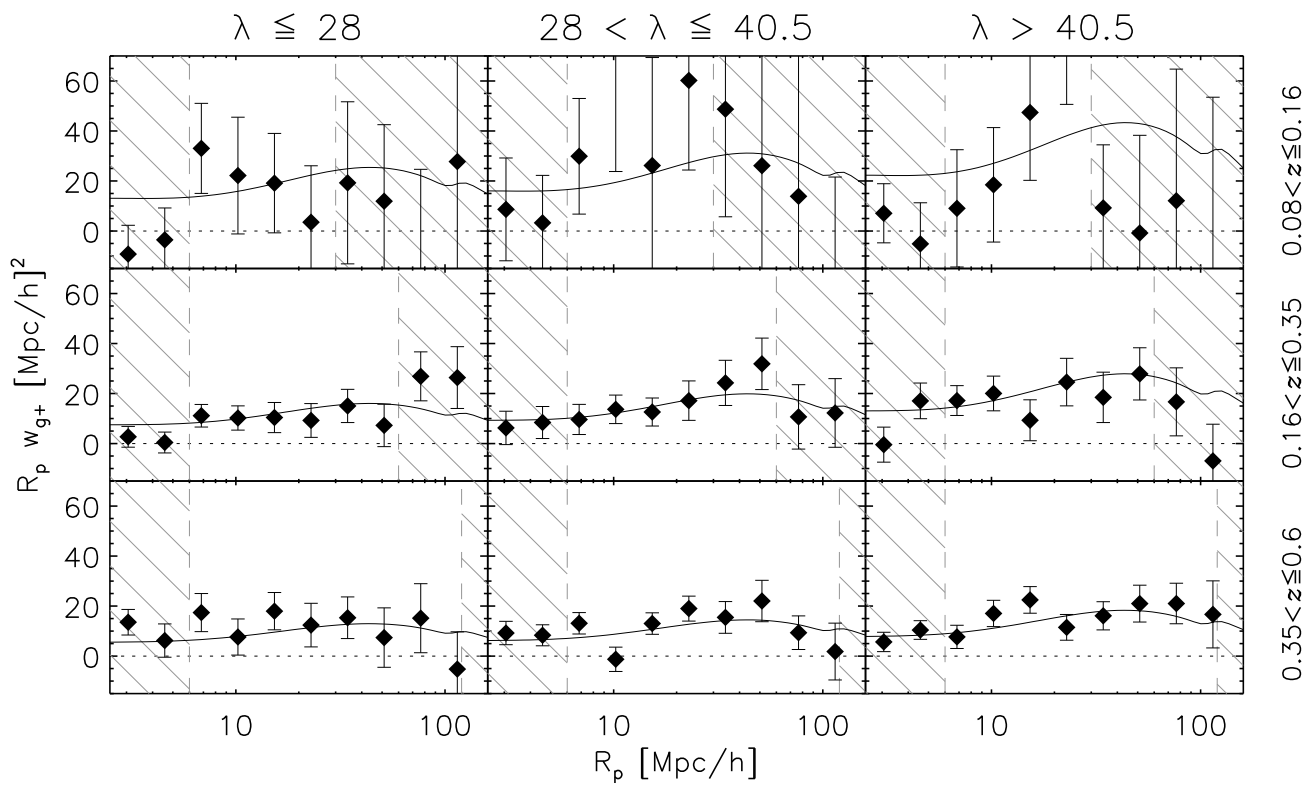}
   \caption{Cluster shape - density correlation of redMaPPer clusters as a function of transverse distance. Different columns correspond to different richness cuts, while different rows correspond to different redshift cuts. We find a positive detection in each panel, meaning that clusters point towards neighbouring clusters. The hashed regions are excluded from the fits, as detailed in the text. The solid black lines are the best-fitting models from the simultaneous fit of the IA model to all the measurements.}
   \label{plot_clusIA}
\end{figure*}
\indent In Fig. \ref{plot_lamz}, we plot the distribution of cluster richness $\lambda$ versus redshift. We split the sample into three redshift bins, and each redshift bin is further subdivided into three richness subsamples. The number of clusters of each richness subsample, as well as the mean redshift, richness and ellipticity, are listed in Table \ref{tab_clusprop}. The shapes of the clusters in each richness subsample are correlated with the positions of all clusters in the same redshift range. The resulting nine cluster shape - density correlation signals enable us to study potential trends with richness and redshift (i.e. constrain $\eta$ and $\beta$ in Eq. \ref{eq_pi2}). We note that our highest redshift bin is incomplete, particularly towards low richnesses. This is partly accounted for by folding in the actual redshift distribution in the model. A remaining limitation is that we use the mean redshift and richness of each subsample in Eq. (\ref{eq_pi2}), ignoring potential higher order moments. Given the current accuracy of our data, we expect this assumption to be harmless. \\
\begin{figure}
   \centering
   \includegraphics[width=1\linewidth]{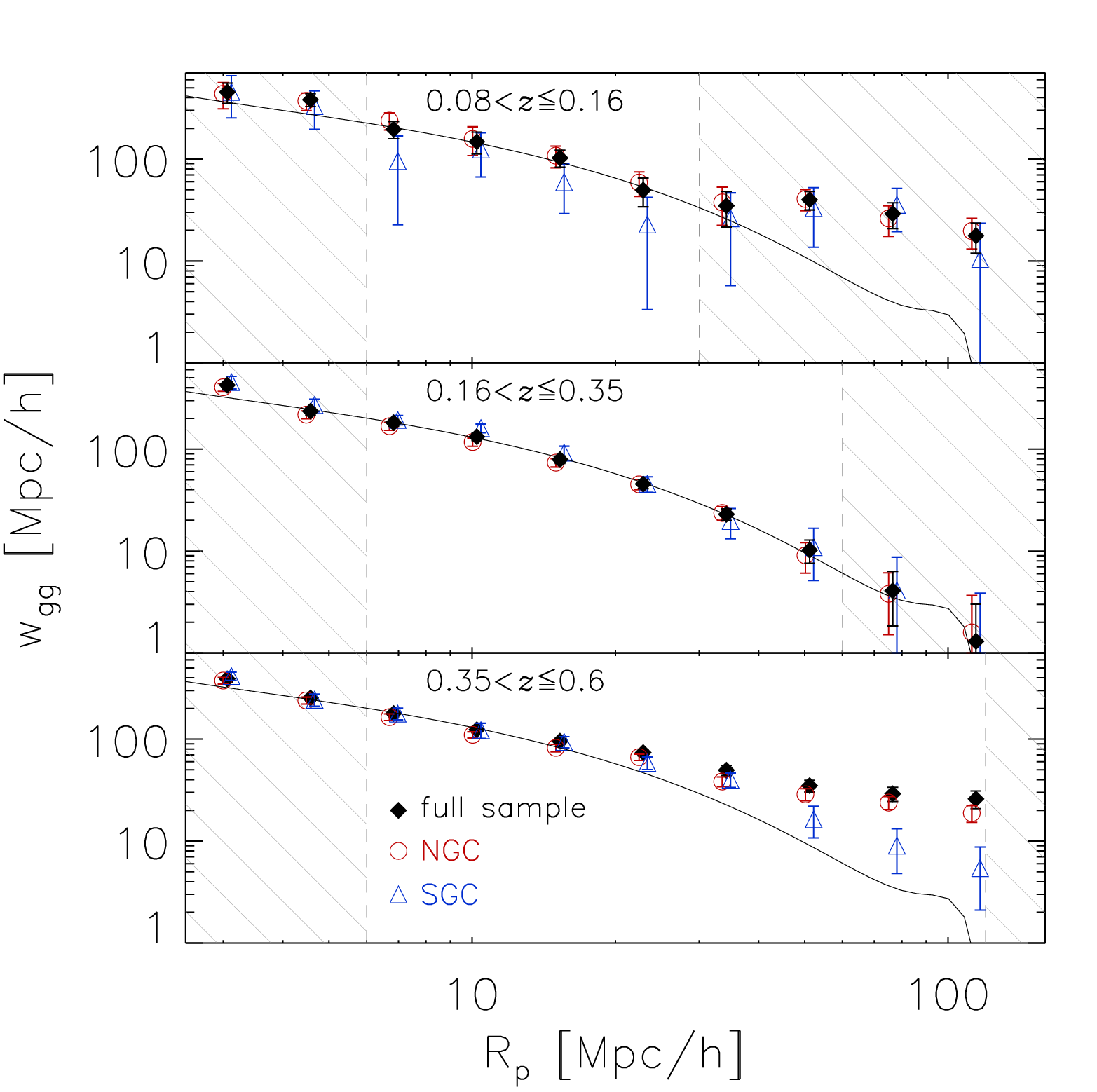}
   \caption{Clustering signal of redMaPPer clusters as a function of transverse separation. Each panel corresponds to a different redshift slice. Solid diamonds show the clustering signal obtained using the total area, while the blue open triangles and red open circles show the signal in the NGC and SGC patch, respectively. The solid lines in each panel show the best-fitting model, fit to each measurement separately. For the highest redshift slice, we fitted the clustering signal of the SGC instead of the full sample, as the signal in the NGC is systematically higher at large scales, presumably due to systematics. The hashed regions are excluded from the fits.}
   \label{plot_clustering}
\end{figure}
\indent To measure the cluster shape - density correlation, we adopted $\Pi_{\rm max}=-\Pi_{\rm min}=$ 100 Mpc/$h$, 125 Mpc/$h$ and 150 Mpc/$h$ in Eq. (\ref{eq_wg+m}) and (\ref{eq_wggm}) for our low-redshift, intermediate-redshift and high-redshift sample, respectively. The increase in range accounts for the increase of photometric redshift scatter of redMaPPer clusters from $\sigma_z\approx0.006$ to $\sigma_z\approx0.02$ between $z\approx0.1$ and $z\approx0.5$, which spreads the signal over larger ranges in $\Pi$. To test whether these integral ranges encapsulate all the signal, we extended the line-of-sight range by 25 Mpc/$h$ on a side and found that this did not significantly increase the signals. In practise, we replaced the integrals of Eq. (\ref{eq_wg+m}) and (\ref{eq_wggm}) by a sum over 20 line-of-sight bins, each having a width of $(\Pi_{\rm max}-\Pi_{\rm min})/20$ Mpc/$h$. \\
\indent We measured the signal as a function of transverse comoving separation in 10 logarithmically-spaced bins between 2.5 Mpc/$h$ and 140 Mpc/$h$. To determine the covariance matrix, we used a jackknife technique. We defined 45 non-overlapping jackknife patches, rectangles of approximately 16$\times$16 degrees, over the entire survey area. $\sim$5\% of the redMaPPer clusters could not be covered by our rectangles and were excluded from the analysis. The inverse covariance matrix that we used in the fit was corrected for a bias that is introduced when noisy covariance matrices are inverted \citep{Kaufmann67,Hartlap07}. \\
\indent We restricted the analysis to mildly non-linear scales of $R_{\rm p}>6$ Mpc/$h$. The upper scale in the fit was set by the lower redshift cut of each cluster sample. At $z=0.08$, $z=0.16$ and $z=0.35$, 8 degrees (i.e. half the size of a jackknife patch) roughly corresponds to 30 Mpc/$h$, 60 Mpc/$h$ and 120 Mpc/$h$, which we adopted as the maximum scale in the fit. The jackknife errors at larger transverse distances become increasingly less representative of the true measurement errors. \\
\indent We used the random catalogue from redMaPPer to measure the correlations with random points ($S_+R$, $D_{\rm s} R$, $D_{\rm d} R$ and $RR$). The random catalogue is $\sim$100 times denser than the real cluster catalogue. To speed up our calculations, we downsampled the random catalogue to an overdensity of $\sim$5, which is sufficient for our purposes. \\


\section{Results}\label{sec_res}
We show the cluster shape - density correlation in Fig. \ref{plot_clusIA}. At low redshift, the error bars are large as the number of clusters is small, but we obtain a tentative detection of a positive alignment. At intermediate and high redshift, we obtain a clear detection in all our richness subsamples. Detecting a positive signal means that clusters point towards neighbouring clusters. \\
\indent The projected clustering signal of redMaPPer clusters is shown in Fig. \ref{plot_clustering}. Neighbouring radial bins are correlated, particularly at large scales. The clustering signal of these clusters was also measured in \citet{Baxter16}, in order to constrain the mass-richness relation. As they reported differences in the clustering signal in the North and South Galactic Cap (NGC and SGC, respectively), we also analysed them separately, switching to jackknife patches of 8$\times8$ deg instead of our nominal 16$\times16$ deg to guarantee a sufficient number of jackknife realizations to estimate the covariance matrices. The clustering signals of our $0.08<z\leq0.16$ and $0.16<z\leq0.35$ bins broadly agree, but at $0.35<z\leq0.6$, they differ noticeably at transverse separations $R_{\rm p}>30$ Mpc/$h$. The cluster sample is incomplete at this redshift range, making it more susceptible to spatially varying systematics that bias the clustering signal high. Since the signal is comparable at small scales, but systematically higher in the NGC at large scales (in fact consistent with a constant additive bias), we expect that those measurements are affected by systematics. A systematic in the spatial distribution of galaxy clusters is not expected to affect the $w_{\rm g+}$ measurements, as removing or adding cluster shape - density pairs does not affect the mean signal, only its error.\\
\indent To obtain the bias of the three density samples, we fitted Eq. (\ref{eq_wgg}) to the clustering signal of the full samples. For the low- and intermediate-redshift sample, we obtained good fits with corresponding reduced $\chi^2$ values of 0.21 and 0.25, respectively, suggesting that we may have overestimated the error bars somewhat. For the high-redshift sample, we obtained an unacceptably high $\chi^2$, likely because of the systematics. Fitting the signal of the SGC patch, however, led to an acceptable fit with $\chi_{\rm red}^2=1.56$. Therefore, we decided to use the clustering signal measured in the SGC in order to determine the bias of our high-redshift sample. To ensure that the apparent mismatch between model and data on large scales does not bias our results, we repeated the fits on scales $R_{\rm p}<20$ Mpc/$h$ only. The resulting constraints on the cluster bias were virtually unchanged. The reason is that the fit is driven by the innermost radial bins, whose errors are smallest, and because the points are correlated. The constraints on the biases, which are listed in Table \ref{tab_clusprop}, are in reasonable agreement with the results from \citet{Baxter16}, who reported biases in the range 3 -- 5 for redMaPPer clusters at $\lambda>20$ and $0.1<z<0.33$. \\
\begin{figure}
   \centering
   \includegraphics[width=0.85\linewidth,angle=270]{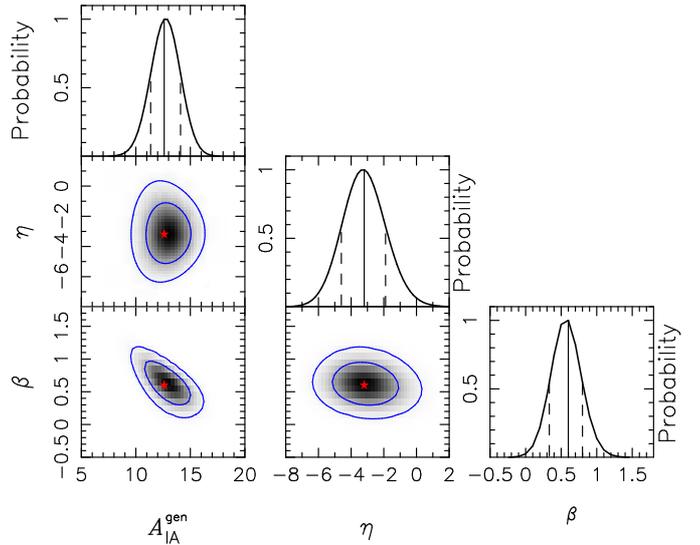}
   \caption{Marginalized 1-D and 2-D posteriors of the parameters of the IA model; $A_{\rm IA}^{\rm gen}$ is the amplitude at the pivot location of $z=0.3$ and $\lambda=30$, $\eta$ describes the redshift dependence and $\beta$ the richness dependence. The top panels indicate the 1-D marginalized posteriors normalized to a peak amplitude of 1, the vertical solid lines indicate the best-fitting values and the vertical dashed lines the 68\% confidence intervals. In the 2-D posterior plots, the red stars indicate the best-fitting values, the blue contours are the 1- and 2-$\sigma$ confidence intervals and the grey scale indicates the value of the posterior.}
   \label{plot_post}
\end{figure}
\indent Having obtained the biases of our density samples, we proceeded with constraining the amplitude of the IA model. We fitted the signal in two ways. First, we fitted all measurements simultaneously using the IA model with a redshift and richness dependence (Eq. \ref{eq_pi2}). The best-fitting models of the combined fit are shown in Fig. \ref{plot_clusIA} and they describe the data well. The reduced $\chi^2$ of the fit is $31.2/(45-3)=0.74$. The 1-D and 2-D marginalized posteriors of the fit parameters are shown in Fig. \ref{plot_post}. The overall amplitude is $A_{\rm IA}^{\rm gen}=12.6^{+1.5}_{-1.2}$, the slope of the redshift dependence is $\eta=-3.20^{+1.31}_{-1.40}$ and the slope of the richness dependence is $\beta=0.60^{+0.20}_{-0.27}$, where the errors correspond to the 68\% confidence intervals, obtained by marginalizing over the other parameters. Hence we obtain tentative evidence that the amplitude of the IA signal increases with richness and towards lower redshift (at 2.2$\sigma$ and 2.4$\sigma$, respectively, assuming that the likelihood is Gaussian). \\ 
\indent  We also determined the amplitude of the IA model for each shape sample separately. Most of the reduced $\chi^2$ values of the best-fitting models are between 0.25 and 1, again suggesting we may have overestimated our errors somewhat. The corresponding amplitudes can be found in Table \ref{tab_clusprop}.\\
\begin{figure}
   \centering
   \includegraphics[width=1\linewidth,angle=270]{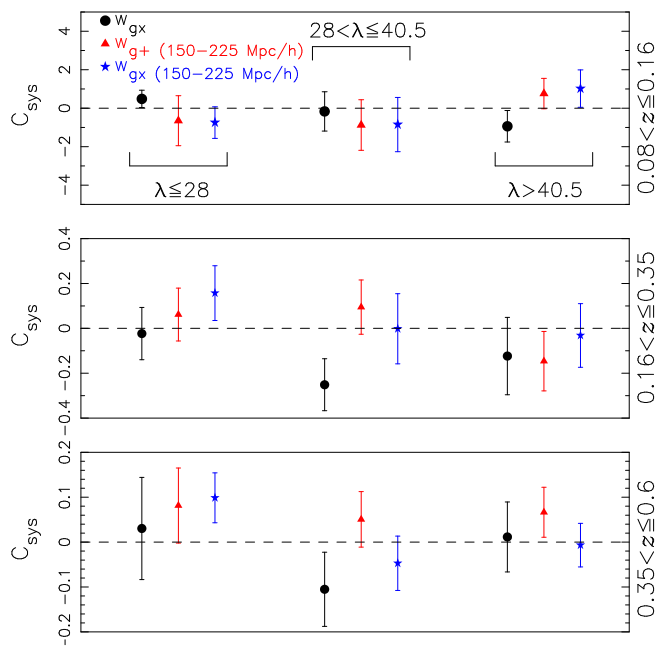}
   \caption{Amplitude of the systematic signal for the three systematic tests we performed, determined using the same scales as in the science analysis. Circles indicate the results of the cross shear - density correlation, triangles indicate the shape - density correlation for line-of-sight separations of $150<\Pi<225$ Mpc/$h$ and stars indicate the cross shear - density correlation for line-of-sight separations of $150<\Pi<225$ Mpc/$h$. The results are split for the three redshift and three richness slices, as indicated in the figure. The scaling of the vertical axis is different for each panel. A significant non-zero signal may indicate the presence of systematics.}
   \label{plot_wgsys}
\end{figure}

\subsection{Systematic tests}
\indent We performed a number of tests to check for the presence of systematics. First, we measured the correlation between the cross-shear component of the cluster shapes and the density field, $w_{\rm g\times}$. The cross shear measures a net curl of cluster shapes, which would violate parity symmetry and is expected to be zero. A non-zero cross shear would therefore indicate systematics in our cluster shapes. Note that for the systematic tests, we did not subtract the cluster shape - random point correlation, as that might remove a systematic signal, if present. \\ 
\indent We measured $w_{\rm g\times}$ for each richness and redshift bin. To quantify the results, we fitted a constant ${\rm C_{sys}}$ on the same scales that we used to fit $w_{\rm g+}$ (i.e. 6 Mpc/$h$ $<R_{\rm p}<$ 30 Mpc/$h$ for the lowest redshift bin). To do the fit, we inverted the jackknife covariance matrix, using only those radial ranges that correspond to the scales used in the fit. The results are shown in Fig. \ref{plot_wgsys}. The amplitudes deviate by $<2\sigma$ from zero, hence none of them indicate the presence of systematics. \\
\indent Next, we measured the $w_{\rm g+}$ and $w_{\rm g\times}$ correlations for line-of-sight separations of $150<\Pi<225$ Mpc/$h$ and $-150<\Pi<-225$ Mpc/$h$. A non-zero signal could indicate problems with the cluster redshifts and/or the presence of additive systematics, and also tests whether there is significant additional signal on scales $R_{\rm p}>150$ Mpc/$h$. We show the constraints on ${\rm C_{sys}}$ for all richness and redshift bins in Fig. \ref{plot_wgsys}. We note that the error bars do not have the exact same size. This difference is mostly suppressed when we only use the diagonals of the inverted covariance matrix in the fit, which we prefer not to do to keep the test as closely as possible to the science analysis. Hence we conclude that this difference is most likely caused by noise in the covariance matrix. As mentioned earlier, we also measured the clustering and IA signals by increasing the line-of-sight ranges of the integrals by 25 Mpc/$h$ on a side and found that the resulting changes of the model parameters were within their 1$\sigma$ errors and hence insignificant. Furthermore, we tested the impact of using 8$\times$8 jackknife patches to measure the IA signals, which made no significant impact either. 

\subsection{Comparison with cluster IA results}
\begin{figure}
   \centering
   \includegraphics[width=1\linewidth]{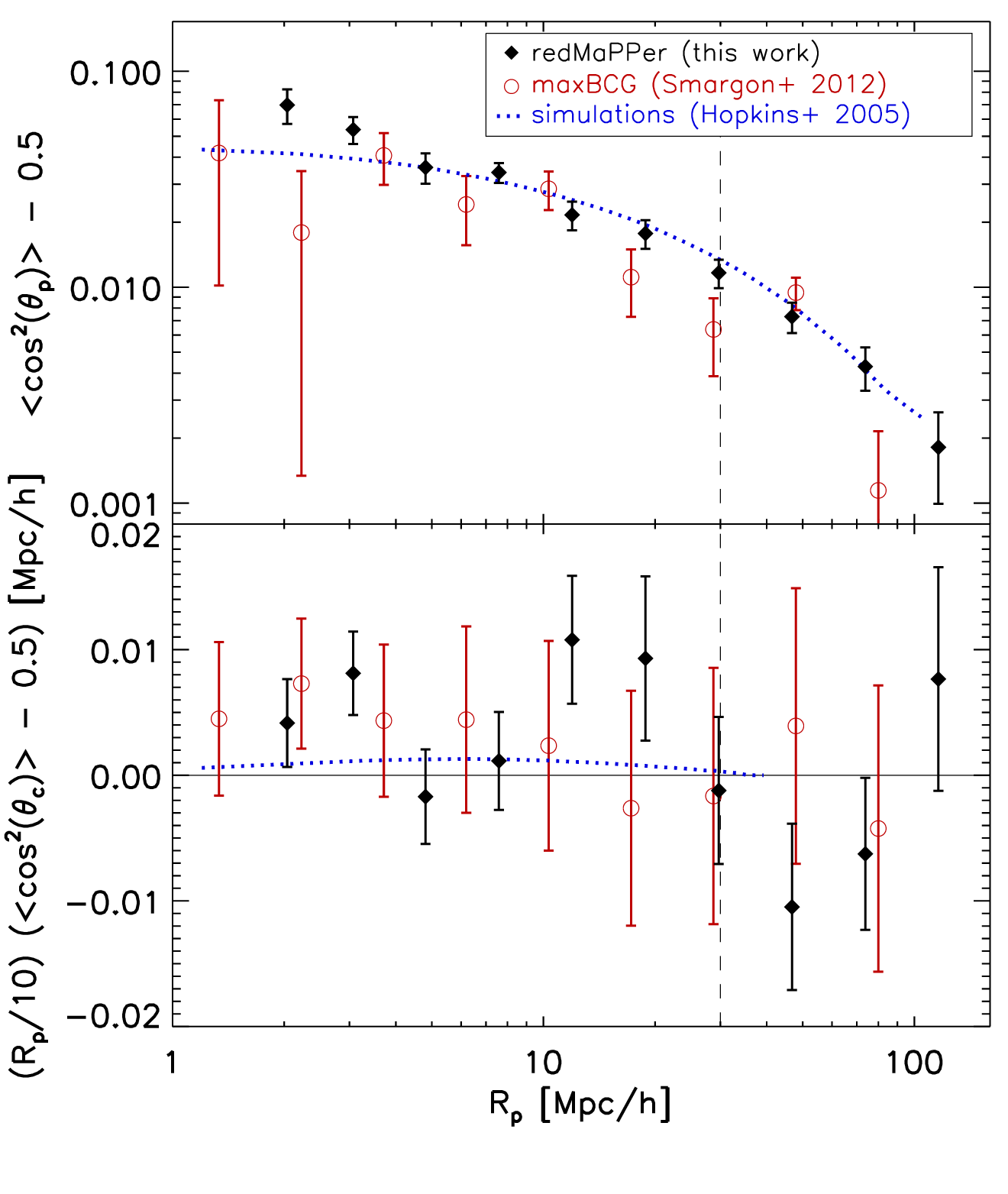}
   \caption{Cluster pointing angle alignment $\langle\cos^2(\theta_p)\rangle$ ({\it top}) and cluster correlation angle alignment $\langle\cos^2(\theta_c)\rangle$ ({\it bottom }) as a function of  transverse separation, determined using all cluster pairs with a line-of-sight separation less than 50 Mpc/$h$. The vertical dashed line indicates half the size of our jackknife patch at the minimum cluster redshift; the measurement errors at larger scales become increasingly less reliable. Our results are indicated by solid black diamonds, the open circles show the measurement for maxBCG clusters from \citet{Smargon12} and the blue dotted line the simulation results for projected cluster shapes from \citet{Hopkins05}, projected along the line of sight.}
   \label{plot_cos2p}
\end{figure}

\indent The most comparable and recent work on the cluster shape - density correlation was presented in \citet{Smargon12}, who studied two cluster samples, the maxBCG catalogue \citep{Koester07} and an adaptive matched filter catalogue \citep{Dong08}, containing 6625 and 8081 clusters at $0.1<z<0.3$ and $0.08<z<0.44$, respectively. We focus on the maxBCG results, as that sample has been more widely studied, and because the results of the two samples are similar. \\
\indent The shape - density correlation was measured with a different statistic, $\langle \cos^2(\theta_p)\rangle$, with $\theta_p$ the angle on the sky between the projected major axis of the cluster and the transverse separation vector of cluster pairs. This estimator has an expectation value of 0.5 for uncorrelated angles, while a value large than 0.5 implies that the major axes of clusters point on average towards neighbouring clusters. To enable a comparison with our results, we measured the signal using the same estimator for the full redMaPPer catalogue (except the clusters outside the jackknife patches). \\
\indent \citet{Smargon12} measured the signal for cluster pairs with a maximum transverse separation of 100 Mpc/$h$, and separated in redshift by less than 0.015, which corresponds to a maximum line-of-sight separation of $\sim$50 Mpc/$h$. To mimic their selection, we stack the signal of all cluster pairs with a line-of-sight separation less than 50 Mpc/$h$ in a given transverse separation bin. The results are shown in Fig. \ref{plot_cos2p}. The measurements agree reasonably well on all scales. Note that our measurement errors at $R_{\rm p}>30$ Mpc/$h$ should be interpreted with care, as these separations are larger than half the size of a jackknife patch at the lowest redMaPPer cluster redshift, $z=0.08$. The measurements themselves should be robust. \\
\indent  \citet{Smargon12} compared their measurements to the results of \citet{Hopkins05}, which are based on fiducial $\Lambda$CDM dark-matter-only simulations with a particle mass of $1.264\times10^{11} M_\odot/h$. Only haloes with more than 160 particles ($M_h>2\times10^{13}M_\odot/h$) were considered, hence their shapes are robustly determined.  The $\langle\cos^2(\theta_p)\rangle$ measured in the data was considerably lower than the one of the simulations, and \citet{Smargon12} argued that this might be due to observational systematics. However, while our measurements and those of \citet{Smargon12} project $\langle\cos^2(\theta_p)\rangle$ along the line of sight, \citet{Hopkins05} measured $\langle\cos^2(\theta_p)\rangle$ as a function of 3-D separation and only projected the ellipticities\footnote{Figure 7 and 9 of \citet{Hopkins05} are labeled with ``2-D Projection", but that only referred to the ellipticities, which caused the confusion.}. Therefore, to enable a fair comparison, we projected the results from \citet{Hopkins05} by averaging them in the range from $R_{\rm p}$ to $\sqrt{R_{\rm p}^2+\Pi^2_{\rm max}}$, with $\Pi_{\rm max}=50$ Mpc/$h$. The results are shown in Fig \ref{plot_cos2p}. The model predictions describe the data well at $R_{\rm p}>4$ Mpc/$h$, particularly considering that it is not a fit. \\
\indent  At $R_{\rm p}<4$ Mpc/$h$, our measurements are higher than the model. A possible cause is contamination; some of the cluster members in the redMaPPer membership catalogues may be interlopers \citep[see e.g.][]{Zu16}. Even if they have a low membership probability, they are not excluded when we estimate the cluster ellipticity and could cause a bias, particularly if these interlopers come from a neighbouring cluster. To estimate whether this effect is large, we determined the cluster ellipticities using cluster members with $p_{\rm mem}>0.5$ only (instead of $p_{\rm mem}>0.2$). Applying this cut shifts the first point down by 1$\sigma$, which makes it consistent with the simulation results. A more conservative cut of $p_{\rm mem}>0.8$ was found necessary in \citet{Zu16} to remove the impact of projection effects altogether (to within the errors). Such a cut is not feasible here, as the decrease in usable satellite galaxies would increase the Poisson noise of the cluster's major axis estimate, diluting the signal on all scales. To make sure projection effects are not important, we implemented another test in which we removed all clusters from the shape sample that have a neighbouring cluster with a transverse separation $<3$ Mpc/$h$ and with a redshift difference smaller than 0.03. This removes clusters whose shape are most likely to be affected by projection effects ($\sim$9\% of the total). The $\langle\cos^2(\theta_p)\rangle$ signal at $R_{\rm p}>3$ Mpc/$h$ is not affected by this cut. \\
\indent A number of observational errors were identified in \citet{Smargon12} that might have suppressed $\langle\cos^2(\theta_p)\rangle$ in the data, such as photometric redshift errors, Poisson noise in estimating the cluster major axis due to the availability of a low number of satellites, cluster centroiding errors and errors in the major axis estimates of clusters whose member distribution is intrinsically nearly round. The impact of the first two of these, that is photometric redshift errors and Poisson noise of the cluster major axis, were estimated to have lowered their measurements by $\sim$20\%. We also tested the impact of a number of systematics; we list them below and describe how we estimated their impact on the measurements. The results of these tests are summarized afterwards.
\begin{itemize}
\item[$\bullet$] photo-$z$ errors: The redMaPPer clusters have a photometric redshift error, which shifts them along the line of sight, diluting the signal. We cannot undo that, but we can estimate whether our results are sensitive to it. To do that, we scatter the cluster redshifts, by adding a number that is randomly drawn from a Gaussian with a standard deviation of 0.01, which is the typical photo-$z$ error of redMaPPer clusters. 
\item[$\bullet$] miscentering: If the cluster centre is not correctly identified, the inferred cluster ellipticity and hence the major axis becomes biased, which reduces $\langle\cos^2(\theta_p)\rangle$. Comparing the BCG location with the X-ray centre from overlapping X-ray catalogues, \citet{Rozo14} estimated that $\sim$14\% of the redMaPPer clusters are miscentred. The cumulative distribution of the miscentring radii of those clusters increases roughly linear to $R_{\rm p}=0.8$ Mpc/$h$, which corresponds well with a Rayleigh distribution with a width of 0.3 Mpc/$h$ \citep{Simet16}. We cannot undo the miscentring in redMaPPer, but we can estimate whether our results are sensitive to it by introducing an additional miscentring. We randomly displaced the cluster centre for 14\% of the clusters, by an amount that is randomly drawn from a Rayleigh distribution with a width of 0.3 Mpc/$h$. We remeasured the ellipticities of these clusters and repeated the measurement with the updated ellipticities and positions.
\item[$\bullet$] major axis errors: To estimate the effect of major axis errors for nearly round clusters, we removed all clusters with $\eps\leq0.05$ and repeated the measurements with the remaining sample. In principle, one should also update the predictions from simulations by applying the same cut. Nonetheless, it gives us an estimate what the potential impact of this effect is.
\item[$\bullet$] merging clusters: Some highly elongated or merging clusters may be split by the redMaPPer cluster finder into two separate parts with different, biased position angles. To test for such an effect, we removed all clusters with a neighbour at $R_{\rm p}<5$ Mpc/$h$ and $\Pi<100$ Mpc/$h$  from the shape sample (but not the density sample). These cuts removed $7\,322$ clusters. 
\item[$\bullet$] completeness: Figure \ref{plot_lamz} shows that the cluster sample becomes increasingly incomplete towards higher redshift. To test whether that has an effect, we repeated the measurement using clusters at $z\leq0.35$, where the sample is nearly complete.
\end{itemize}
We assessed the impact of each test separately. The only systematic that had a large impact on the measurement was photo-$z$ errors, which shifted all points down by 0.5-1$\sigma$. The impact of all other effects was considerably smaller and can be safely ignored in this comparison. A crude correction for the effect of photo-$z$ errors would be to shift our measurements upward by 0.5-1$\sigma$, which would improve the agreement at $R_{\rm p}>10$ Mpc/$h$, but cause an increased overestimation of the signal at smaller scales. For completeness, we note that \citet{Hopkins05} tested whether their results depend on the scales used to measure the ellipticity and reported only a minor effect for the least massive haloes, making it unlikely that the use of different ellipticity estimators has a significant impact on the comparison. \\
\indent Studies of luminous red galaxy (LRG) samples have reported highly significant detections of the IA signal \citep[e.g.][]{Mandelbaum06,Joachimi11,Singh15}. These observations have been compared with results from simulations, which revealed that simulations predict larger signals than what has been observed. To reconcile the two, \citet{Okumura09} proposed the presence of a significant amount of misalignment between the orientations of LRGs and their dark matter haloes. A detailed comparison between observations and simulations is complicated, however, by the fact that the IA signal sensitively depends on the shape measurement method used to measure LRG shapes, with methods giving more weight to larger distances from the LRG's centre producing larger IA signals \citep{Singh16IA}. \\
\indent Similar trends have been reported in hydrodynamical simulations. \citet{Velliscig15} measured the IA signal in the EAGLE simulations \citep{Schaye15} and found that, when all star particles in each halo were used to estimate the shapes of galaxies, the IA signal of a LRG-like sample was overpredicted; however, by only using star particles inside a radius that contained half the stellar mass of the halo, the observations could roughly be matched. \citet{Tenneti15} measured the IA signal of a LRG-like sample in the MassiveBlack-II simulations \citep{Khandai15} and reported good agreement with observations for SDSS LRGs; however, they used a reduced inertia tensor to define the galaxy ellipticities, downweighting particles further away from the centre, effectively similar to what was done in \citet{Velliscig15}. It is not yet clear how the shapes of galaxies in hydrodynamical simulations compare to the shapes of LRGs that are measured in the data. It seems unlikely that these two aforementioned effects affect the cluster IA signal much, as a very misaligned satellite distribution would not be dynamically stable, and because the details of how we measure cluster shapes are unlikely to affect the major axis much. Hence comparing the IA signal of clusters with simulations should be more straightforward, which makes it a powerful probe of large-scale structure. \\
\indent We also measure $\langle\cos^2(\theta_ c)\rangle$, with $\theta_c$ the angle between the projected major axes of a pair of clusters. The measurements are shown in the lower panel of Fig. \ref{plot_cos2p}. As before, we projected the simulation results of \citet{Hopkins05} to enable a comparison. Since $\langle\cos^2(\theta_ c)\rangle$ has much smaller values in the simulation than $\langle\cos^2(\theta_ p)\rangle$, and since it decreases faster with 3-D separation as well, the projected signal is very small, implying that it cannot be detected with current cluster samples, unless photometric redshifts improve or spectroscopic redshifts are used. Our results on $R_{\rm p} < 30$ Mpc/$h$ do seem to suggest the presence of a signal, with an amplitude that is larger than the model, but the errors of our measurements are still fairly large.  \\

\subsection{Trend with halo mass}
\indent The amplitude of the IA signal of LRGs increases with luminosity \citep{Joachimi11,Singh15}. It is interesting to check whether this trend continues with galaxy clusters: if clusters can be viewed as higher luminosity (or mass) extensions of LRGs, the cluster shape - density correlation could be used to improve IA models for galaxies, which would benefit future cosmic shear surveys. \\
\begin{figure}
   \centering
   \includegraphics[width=1\linewidth]{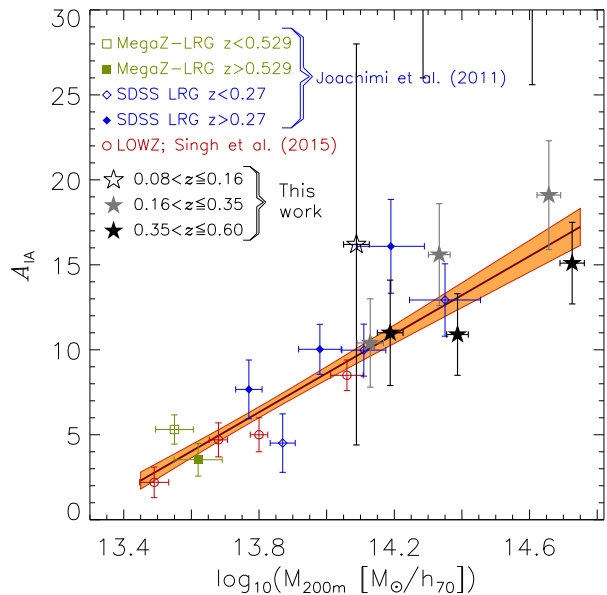}
   \caption{Amplitude of the linear alignment model as a function of halo mass. Our results are indicated by the stars. We also show literature results for luminous red galaxies from \citet{Singh15} and \citet{Joachimi11}; the masses for these samples were determined from their mean luminosities using the luminosity-to-halo mass relation from \citet{VanUitert15}. The solid line indicates the best-fitting linear relation between the $\log_{10}$ of halo mass and the IA amplitude, and the orange contours indicate the 1$\sigma$ model uncertainty of this fit. Two of our redMaPPer low-redshift results fall outside the plotted range, but both of them are within 2$\sigma$ of the best-fitting relation.}
   \label{plot_massAIA}
\end{figure}
\indent In order to compare our results with those obtained for LRGs, we converted both the LRG luminosities and the cluster richnesses to halo mass. All halo masses we derived are defined as the mass inside a sphere where the density exceeds 200 times the mean matter density at $z=0$. To convert cluster richnesses to halo mass, we used the mass-richness relation from \citet{Simet16}, which was derived from a weak lensing analysis of the same redMaPPer cluster sample. Parametrizing this relation as $M=M_0 (\lambda/\lambda_0)^\alpha$, \citet{Simet16} reported $\log_{10}(M_0)=14.344\pm0.031$ (where we added the statistical and systematic errors in quadrature) and $\alpha=1.33^{+0.09}_{-0.10}$ for a pivot richness $\lambda_0=40$, with little correlation between $M_0$ and $\alpha$. We computed the mean halo masses using the mean richnesses listed in Table \ref{tab_clusprop}. We note that \citet{Simet16} only used clusters at $0.1\leq z\leq0.33$, hence the masses of our high redshift sample may be somewhat biased if the mass-richness relation evolves at $z>0.33$. Furthermore, the halo masses from \citet{Simet16} are defined with respect to the mean matter density at the cluster's redshift, and we converted them to our definition, which increased the masses by $\sim$15\%. We show our constraints on $A_{\rm IA}$ of the individual shape samples as a function of halo mass in Fig. \ref{plot_massAIA}. The horizontal error bars correspond to the propagated uncertainties from the mass-richness relation.\\
\indent We compare these constraints with results for LRGs from two studies. \citet{Singh15} measured the IA signal for LOWZ LRGs. To convert their luminosities to halo mass\footnote{\citet{Singh15} also provide mass estimates for their LRG samples, derived from a weak lensing analysis. We did not use them for a number of reasons: they are defined with respect to a different overdensity, they were derived using a different mass-concentration relation, and they did not account for the scatter in the luminosity-to-halo mass relation. Since we needed to convert the luminosities of \citet{Joachimi11} with the scaling relations from \citet{VanUitert15} anyway, we decided to use the same scaling relations to convert the luminosities from \citet{Singh15}, for consistency and for easing a comparison of results.}, we used the luminosity-to-halo mass relation from \citet{VanUitert15}, which was determined for LOWZ and CMASS galaxies \citep{Dawson13} in two separate, non-overlapping redshift bins each (hence four redshift bins in total), covering a redshift range from \mbox{$\langle z \rangle=0.2$} to $\langle z \rangle=0.6$. The relation in each redshift bin was parametrized as \mbox{$M=M_{0,L} (L/L_0)^{\beta_L}$}, with a pivot luminosity $L_0=10^{11}h_{70}^{-2}L_\odot$. We took a weighted mean of the amplitude and slope of the luminosity-to-halo mass relation for the two LOWZ redshift bins (at $0.15<z<0.29$ and $0.29<z<0.43$) from \citet{VanUitert15}, resulting in $M_{0,L}=5.68\pm0.34\times10^{13} h_{70}^{-1}M_\odot$ and $\beta_L=1.50\pm0.20$. We propagated the uncertainties of these power law parameters to uncertainties in halo mass. These results are also shown in Fig. \ref{plot_massAIA}. \\
\indent Next, we compare with the results from \citet{Joachimi11}, who measured the shape - density correlation signal of LRGs from the MegaZ-LRG sample \citep{Collister07} and from the SDSS \citep{Eisenstein01}. \citet{Joachimi11} applied an additional colour cut to the MegaZ-LRG sample to make its colour-magnitude relation resemble the one of SDSS LRGs, for consistency. To convert the luminosities to halo mass, we used the luminosity-to-halo mass relation from \citet{VanUitert15} that was nearest in redshift. For the $z<0.529$ and $z>0.529$ MegaZ-LRG samples, we used the scaling relation for CMASS galaxies at $0.43<z<0.55$ and $0.55<z<0.7$, respectively, while for the $z<0.27$ and $z>0.27$ SDSS-LRG samples, we used the scaling relation for LOWZ galaxies at $0.15<z<0.29$ and $0.29<z<0.43$, respectively. The mean redshift of the LRG samples from \citet{Joachimi11} differs at most by 0.05 from the mean redshift at which the scaling relations from \citet{VanUitert15} were determined, hence we do not expect a significant redshift evolution over such a small redshift range. The results are shown in Fig. \ref{plot_massAIA}. \\
\indent We find a fairly tight relation between halo mass and IA amplitude, with our results for redMaPPer clusters smoothly extending the trend of LRGs. We quantify this relation using 
\begin{equation}
A_{\rm IA} = B_{\rm IA} \log_{10}(M_{200m}/M_{\rm piv}) + C_{\rm IA} \;,
\end{equation}
and fit for $B_{\rm IA}$ and $C_{\rm IA}$, adopting $M_{\rm piv}=10^{13.8} M_\odot/h$ as that nearly decorrelates the fit parameters. In the fit, we only account for the errors on $A_{\rm IA}$, because the errors on the mass are smaller than the errors on the IA amplitude. We separately assess the impact of the mass errors below (which is complicated by the fact that the masses that were derived by the same scaling relation are fully correlated). We obtain $B_{\rm IA}=11.5\pm1.1$ and $C_{\rm IA}=6.3\pm0.3$. The reduced $\chi^2$ of the best-fitting model is $26.4/(21-2)=1.4$, hence this model provides a reasonable description of this combined data set. Excluding our $0.08<z\leq0.16$ results from the fit leads to an improved reduced $\chi^2$ of 1.2, without significantly changing the fit parameters. \\
\indent To estimate the impact of the mass errors, we used Monte Carlo simulations in which we reassigned the masses of the samples by scattering the scaling relation parameters which we used to convert richness and luminosity to halo mass, and repeating the fit between IA amplitude and (scattered) mass. This procedure, which preserves the correlation in the masses, was repeated $10\;000$ times, and the spread in the best-fitting values of $B_{\rm IA}$ and $C_{\rm IA}$ was taken to be the error caused by the errors on mass. Combining the thus obtained errors with the ones quoted above (i.e. assuming they are completely independent), we found that the errors increased by $\sim$20\%, a fairly small change but not entirely negligible. Finally, we note that the small differences in the adopted values of $\Omega_{\rm M}$ and $\sigma_8$ in the works we compared to are not expected to have a significant impact on the comparison. \\
\indent This smooth and continuous trend of IA amplitude from LRGs to galaxy clusters is somewhat surprising. Cluster satellites are expected to trace the overall distribution of dark matter, while the orientation of LRGs traces the matter distribution deep inside the halo. Various studies have reported the presence of ellipticity gradients in dark matter haloes. Haloes become rounder towards larger scales \citep[e.g.][]{Allgood06,Despali16} and their orientations change as well. For example, the mean misalignment angle between the dark matter distribution at the halo's centre and at the virial radius is $\sim$20 degrees for LRG-sized haloes in the dark-matter-only simulations of \citet{Despali16}, while \citet{Wang14} and \citet{Velliscig15} report a mean 3-D misalignment angle between central galaxies and their haloes within the virial radius of $\sim$35 degrees using hydrodynamical simulations; the misalignment angle decreases for more massive haloes but does not become zero. These values are in line with the results from \citet{Huang16}, who found mean misalignment angles between the position angle of redMaPPer BCGs and their satellite distributions of 32--35 degrees, depending on which shape measurement technique was employed. Further evidence of this scenario comes from a halo ellipticity study of galaxy groups \citep{VanUitert16}, in which it was found that the measured BCG ellipticity traces the projected mass distribution at scales $<0.5\times r_{200}$, while the projected distribution of satellites traces the projected mass distribution at the virial radius. \\
\indent A misalignment between the LRGs and their haloes reduces the IA signal of LRGs, and one would expect the IA signal of LRGs to be systematically lower than those of clusters. However, haloes become rounder at large scales; the average ellipticity of redMaPPer clusters is $\sim$0.12 (see Table \ref{tab_clusprop}), a factor of $\sim$2 smaller than what is typically measured for the BCGs \citep{Velliscig15,Huang16}. This causes a similar reduction of the IA amplitude. The smooth and continuous trend that we find thus suggests that the ellipticity component projected towards the density peaks is similar for the central part of haloes (traced by the LRGs) and for the haloes as a whole (traced by the satellites). It would be interesting to check whether a similar trend is observed in large N-body, or ideally, hydrodynamical simulations. In particular, such simulations could address whether the trend we observe is purely a physical effect, or whether it is partly shaped by systematics, such as measurement noise, selection effects of the redMaPPer cluster finder (i.e. redMaPPer detects clusters in circular apertures, which might bias their ellipticities low), and interlopers.


\section{Conclusions}\label{sec_conc}

We measured the correlation between cluster shapes and the density field using the redMaPPer cluster catalogue in the SDSS, which contains 26 111 clusters and over 1.7 million candidate cluster members. The cluster shapes were estimated using the projected distribution of cluster members, while the density field was traced by the spatial distribution of the same clusters. We separately analysed clusters at low, intermediate and high redshift, and split each redshift slice in three richness subsamples, enabling us to distinguish potential trends with cluster redshift and richness. \\
\indent We detected a positive alignment in all cluster shape samples, showing that clusters point on average towards neighbouring clusters. To interpret the data, we first determined the bias by fitting a model to the clustering signal of redMaPPer clusters. For our low-and intermediate-redshift samples we obtained good fits, but for our high-redshift sample we noticed an excess of clustering signal at large scales in the NGC patch compared to the SGC patch, indicative of a systematic. Therefore, we used the clustering signal in the SGC for this redshift bin. Fixing the thus obtained biases to their best-fitting values, we fitted our $w_{\rm g+}$ measurements using a linear alignment model, explicitly allowing for a redshift and a richness dependence (Eq. \ref{eq_pi2}). We constrained the amplitude at the pivot redshift $z=0.3$ and pivot richness $\lambda=30$ to $A_{\rm IA}^{\rm gen}=12.6^{+1.5}_{-1.2}$ The slope of the redshift dependence is given by $\eta=-3.20^{+1.31}_{-1.40}$ and the slope of the richness dependence is $\beta=0.60^{+0.20}_{-0.27}$, hence we obtained tentative evidence of an increase in the IA signal towards higher richness and towards lower redshift. \\
\indent Our measurements agree well with an earlier study of maxBCG clusters \citep{Smargon12}. We identified the source of the tension with the N-body simulation results from \citet{Hopkins05} as a previously misidentified projection effect. After accounting for this, the simulation results and our measurements agree. \\
\indent We compared our measurements to constraints obtained for LRGs and found that the amplitude of the IA model increases smoothly and monotonically with halo mass, from low-mass LRGs up to massive galaxy clusters. A relation that is linear in $\log_{10}$ of halo mass and IA amplitude provides a satisfactory fit for over more than an order of magnitude in halo mass. This agreement is surprising, as LRGs trace the dark matter distribution at small scales, which is misaligned with the overall dark matter distribution. However, clusters are rounder than LRGs, which causes a similar reduction of the cluster IA signal. Our results suggest that the ellipticity component projected towards density peaks is similar for LRGs and clusters. As the cluster IA signal should be less affected by misalignments and depends less on the details of the shape measurement technique, it is likely a more pure probe of the alignment of haloes with the tidal field. Cluster IA is therefore a great complementary probe to derive precise physical models for galaxy IA, to the benefit of the cosmological exploitation of upcoming lensing surveys such as Euclid and LSST. \\
\indent As an aside, we note that our results could be used to improve cluster finding algorithms. For a given cluster, it is more likely to find another one along its major axis than along its minor axis. If two physically close clusters do not point to each other, it is more likely that one of them is a false detection, compared to when they do point to each other.

\paragraph*{Acknowledgements}
We thank the referee for valuable comments which improved the draft. We thank Eric Baxter and Eli Rykoff for making an updated version of the cluster random catalogue available to us,  Rachel Mandelbaum and Phil Hopkins for their quick replies to our inquiries about the former tension between observations and N-body results, and Rachel Mandelbaum in addition for sending us the \citet{Smargon12} results and for valuable feedback on the draft. EvU acknowledges support from an STFC Ernest Rutherford Research Grant, grant reference ST/L00285X/1. BJ acknowledges support by an STFC Ernest Rutherford Fellowship, grant reference ST/J004421/1. \\
\bibliographystyle{mnras}

\begin{thebibliography}{}
\makeatletter
\relax
\def\mn@urlcharsother{\let\do\@makeother \do\$\do\&\do\#\do\^\do\_\do\%\do\~}
\def\mn@doi{\begingroup\mn@urlcharsother \@ifnextchar [ {\mn@doi@}
  {\mn@doi@[]}}
\def\mn@doi@[#1]#2{\def\@tempa{#1}\ifx\@tempa\@empty \href
  {http://dx.doi.org/#2} {doi:#2}\else \href {http://dx.doi.org/#2} {#1}\fi
  \endgroup}
\def\mn@eprint#1#2{\mn@eprint@#1:#2::\@nil}
\def\mn@eprint@arXiv#1{\href {http://arxiv.org/abs/#1} {{\tt arXiv:#1}}}
\def\mn@eprint@dblp#1{\href {http://dblp.uni-trier.de/rec/bibtex/#1.xml}
  {dblp:#1}}
\def\mn@eprint@#1:#2:#3:#4\@nil{\def\@tempa {#1}\def\@tempb {#2}\def\@tempc
  {#3}\ifx \@tempc \@empty \let \@tempc \@tempb \let \@tempb \@tempa \fi \ifx
  \@tempb \@empty \def\@tempb {arXiv}\fi \@ifundefined
  {mn@eprint@\@tempb}{\@tempb:\@tempc}{\expandafter \expandafter \csname
  mn@eprint@\@tempb\endcsname \expandafter{\@tempc}}}

\bibitem[\protect\citeauthoryear{{Agustsson} \& {Brainerd}}{{Agustsson} \&
  {Brainerd}}{2010}]{Agustsson10}
{Agustsson} I.,  {Brainerd} T.~G.,  2010, \mn@doi [\apj]
  {10.1088/0004-637X/709/2/1321}, \href
  {http://adsabs.harvard.edu/abs/2010ApJ...709.1321A} {709, 1321}

\bibitem[\protect\citeauthoryear{{Aihara} et~al.,}{{Aihara}
  et~al.}{2011}]{Aihara11}
{Aihara} H.,  et~al., 2011, \mn@doi [\apjs] {10.1088/0067-0049/193/2/29}, \href
  {http://adsabs.harvard.edu/abs/2011ApJS..193...29A} {193, 29}

\bibitem[\protect\citeauthoryear{{Allgood}, {Flores}, {Primack}, {Kravtsov},
  {Wechsler}, {Faltenbacher}  \& {Bullock}}{{Allgood} et~al.}{2006}]{Allgood06}
{Allgood} B.,  {Flores} R.~A.,  {Primack} J.~R.,  {Kravtsov} A.~V.,  {Wechsler}
  R.~H.,  {Faltenbacher} A.,   {Bullock} J.~S.,  2006, \mn@doi [\mnras]
  {10.1111/j.1365-2966.2006.10094.x}, \href
  {http://adsabs.harvard.edu/abs/2006MNRAS.367.1781A} {367, 1781}

\bibitem[\protect\citeauthoryear{{Baldauf}, {Smith}, {Seljak}  \&
  {Mandelbaum}}{{Baldauf} et~al.}{2010}]{Baldauf10}
{Baldauf} T.,  {Smith} R.~E.,  {Seljak} U.,   {Mandelbaum} R.,  2010, \mn@doi
  [\prd] {10.1103/PhysRevD.81.063531}, \href
  {http://adsabs.harvard.edu/abs/2010PhRvD..81f3531B} {81, 063531}

\bibitem[\protect\citeauthoryear{{Baxter}, {Rozo}, {Jain}, {Rykoff}  \&
  {Wechsler}}{{Baxter} et~al.}{2016}]{Baxter16}
{Baxter} E.~J.,  {Rozo} E.,  {Jain} B.,  {Rykoff} E.,   {Wechsler} R.~H.,
  2016, \mn@doi [\mnras] {10.1093/mnras/stw1939}, \href
  {http://adsabs.harvard.edu/abs/2016MNRAS.463..205B} {463, 205}

\bibitem[\protect\citeauthoryear{{Blazek}, {McQuinn}  \& {Seljak}}{{Blazek}
  et~al.}{2011}]{Blazek11}
{Blazek} J.,  {McQuinn} M.,   {Seljak} U.,  2011, \mn@doi [\jcap]
  {10.1088/1475-7516/2011/05/010}, \href
  {http://adsabs.harvard.edu/abs/2011JCAP...05..010B} {5, 010}

\bibitem[\protect\citeauthoryear{{Blazek}, {Vlah}  \& {Seljak}}{{Blazek}
  et~al.}{2015}]{Blazek15}
{Blazek} J.,  {Vlah} Z.,   {Seljak} U.,  2015, \mn@doi [\jcap]
  {10.1088/1475-7516/2015/08/015}, \href
  {http://adsabs.harvard.edu/abs/2015JCAP...08..015B} {8, 015}

\bibitem[\protect\citeauthoryear{{Bridle} \& {King}}{{Bridle} \&
  {King}}{2007}]{Bridle07}
{Bridle} S.,  {King} L.,  2007, \mn@doi [New Journal of Physics]
  {10.1088/1367-2630/9/12/444}, \href
  {http://adsabs.harvard.edu/abs/2007NJPh....9..444B} {9, 444}

\bibitem[\protect\citeauthoryear{{Catelan}, {Kamionkowski}  \&
  {Blandford}}{{Catelan} et~al.}{2001}]{Catelan01}
{Catelan} P.,  {Kamionkowski} M.,   {Blandford} R.~D.,  2001, \mn@doi [\mnras]
  {10.1046/j.1365-8711.2001.04105.x}, \href
  {http://adsabs.harvard.edu/abs/2001MNRAS.320L...7C} {320, L7}

\bibitem[\protect\citeauthoryear{{Chisari} et~al.,}{{Chisari}
  et~al.}{2015}]{Chisari15}
{Chisari} N.,  et~al., 2015, \mn@doi [\mnras] {10.1093/mnras/stv2154}, \href
  {http://adsabs.harvard.edu/abs/2015MNRAS.454.2736C} {454, 2736}

\bibitem[\protect\citeauthoryear{{Chisari} et~al.,}{{Chisari}
  et~al.}{2016}]{Chisari16}
{Chisari} N.,  et~al., 2016, \mn@doi [\mnras] {10.1093/mnras/stw1409}, \href
  {http://adsabs.harvard.edu/abs/2016MNRAS.461.2702C} {461, 2702}

\bibitem[\protect\citeauthoryear{{Codis} et~al.,}{{Codis}
  et~al.}{2015}]{Codis15}
{Codis} S.,  et~al., 2015, \mn@doi [\mnras] {10.1093/mnras/stv231}, \href
  {http://adsabs.harvard.edu/abs/2015MNRAS.448.3391C} {448, 3391}

\bibitem[\protect\citeauthoryear{{Collister} et~al.,}{{Collister}
  et~al.}{2007}]{Collister07}
{Collister} A.,  et~al., 2007, \mn@doi [\mnras]
  {10.1111/j.1365-2966.2006.11305.x}, \href
  {http://adsabs.harvard.edu/abs/2007MNRAS.375...68C} {375, 68}

\bibitem[\protect\citeauthoryear{{Croft} \& {Metzler}}{{Croft} \&
  {Metzler}}{2000}]{Croft00}
{Croft} R.~A.~C.,  {Metzler} C.~A.,  2000, \mn@doi [\apj] {10.1086/317856},
  \href {http://adsabs.harvard.edu/abs/2000ApJ...545..561C} {545, 561}

\bibitem[\protect\citeauthoryear{{Dawson} et~al.,}{{Dawson}
  et~al.}{2013}]{Dawson13}
{Dawson} K.~S.,  et~al., 2013, \mn@doi [\aj] {10.1088/0004-6256/145/1/10},
  \href {http://adsabs.harvard.edu/abs/2013AJ....145...10D} {145, 10}

\bibitem[\protect\citeauthoryear{{Despali}, {Giocoli}, {Bonamigo}, {Limousin}
  \& {Tormen}}{{Despali} et~al.}{2016}]{Despali16}
{Despali} G.,  {Giocoli} C.,  {Bonamigo} M.,  {Limousin} M.,   {Tormen} G.,
  2016, preprint, \href {http://adsabs.harvard.edu/abs/2016arXiv160504319D} {}
  (\mn@eprint {arXiv} {1605.04319})

\bibitem[\protect\citeauthoryear{{Dong}, {Pierpaoli}, {Gunn}  \&
  {Wechsler}}{{Dong} et~al.}{2008}]{Dong08}
{Dong} F.,  {Pierpaoli} E.,  {Gunn} J.~E.,   {Wechsler} R.~H.,  2008, \mn@doi
  [\apj] {10.1086/522490}, \href
  {http://adsabs.harvard.edu/abs/2008ApJ...676..868D} {676, 868}

\bibitem[\protect\citeauthoryear{{Dong}, {Lin}, {Kang}, {Ocean Wang}, {Dutton}
  \& {Macci{\`o}}}{{Dong} et~al.}{2014}]{Dong14}
{Dong} X.~C.,  {Lin} W.~P.,  {Kang} X.,  {Ocean Wang} Y.,  {Dutton} A.~A.,
  {Macci{\`o}} A.~V.,  2014, \mn@doi [\apjl] {10.1088/2041-8205/791/2/L33},
  \href {http://adsabs.harvard.edu/abs/2014ApJ...791L..33D} {791, L33}

\bibitem[\protect\citeauthoryear{{Eisenstein} \& {Hu}}{{Eisenstein} \&
  {Hu}}{1998}]{Eisenstein98}
{Eisenstein} D.~J.,  {Hu} W.,  1998, \mn@doi [\apj] {10.1086/305424}, \href
  {http://adsabs.harvard.edu/abs/1998ApJ...496..605E} {496, 605}

\bibitem[\protect\citeauthoryear{{Eisenstein} et~al.,}{{Eisenstein}
  et~al.}{2001}]{Eisenstein01}
{Eisenstein} D.~J.,  et~al., 2001, \mn@doi [\aj] {10.1086/323717}, \href
  {http://adsabs.harvard.edu/abs/2001AJ....122.2267E} {122, 2267}

\bibitem[\protect\citeauthoryear{{Evans} \& {Bridle}}{{Evans} \&
  {Bridle}}{2009}]{Evans09}
{Evans} A.~K.~D.,  {Bridle} S.,  2009, \mn@doi [\apj]
  {10.1088/0004-637X/695/2/1446}, \href
  {http://adsabs.harvard.edu/abs/2009ApJ...695.1446E} {695, 1446}

\bibitem[\protect\citeauthoryear{{Hartlap}, {Simon}  \& {Schneider}}{{Hartlap}
  et~al.}{2007}]{Hartlap07}
{Hartlap} J.,  {Simon} P.,   {Schneider} P.,  2007, \mn@doi [\aap]
  {10.1051/0004-6361:20066170}, \href
  {http://adsabs.harvard.edu/abs/2007A%26A...464..399H} {464, 399}

\bibitem[\protect\citeauthoryear{{Hilbert}, {Xu}, {Schneider}, {Springel},
  {Vogelsberger}  \& {Hernquist}}{{Hilbert} et~al.}{2016}]{Hilbert16}
{Hilbert} S.,  {Xu} D.,  {Schneider} P.,  {Springel} V.,  {Vogelsberger} M.,
  {Hernquist} L.,  2016, preprint, \href
  {http://adsabs.harvard.edu/abs/2016arXiv160603216H} {} (\mn@eprint {arXiv}
  {1606.03216})

\bibitem[\protect\citeauthoryear{{Hinshaw} et~al.,}{{Hinshaw}
  et~al.}{2013}]{Hinshaw13}
{Hinshaw} G.,  et~al., 2013, \mn@doi [\apjs] {10.1088/0067-0049/208/2/19},
  \href {http://adsabs.harvard.edu/abs/2013ApJS..208...19H} {208, 19}

\bibitem[\protect\citeauthoryear{{Hirata} \& {Seljak}}{{Hirata} \&
  {Seljak}}{2004}]{Hirata04}
{Hirata} C.~M.,  {Seljak} U.,  2004, \mn@doi [\prd]
  {10.1103/PhysRevD.70.063526}, \href
  {http://adsabs.harvard.edu/abs/2004PhRvD..70f3526H} {70, 063526}

\bibitem[\protect\citeauthoryear{{Hirata} \& {Seljak}}{{Hirata} \&
  {Seljak}}{2010}]{Hirata10}
{Hirata} C.~M.,  {Seljak} U.,  2010, \mn@doi [\prd]
  {10.1103/PhysRevD.82.049901}, \href
  {http://adsabs.harvard.edu/abs/2010PhRvD..82d9901H} {82, 049901}

\bibitem[\protect\citeauthoryear{{Hirata}, {Mandelbaum}, {Ishak}, {Seljak},
  {Nichol}, {Pimbblet}, {Ross}  \& {Wake}}{{Hirata} et~al.}{2007}]{Hirata07}
{Hirata} C.~M.,  {Mandelbaum} R.,  {Ishak} M.,  {Seljak} U.,  {Nichol} R.,
  {Pimbblet} K.~A.,  {Ross} N.~P.,   {Wake} D.,  2007, \mn@doi [\mnras]
  {10.1111/j.1365-2966.2007.12312.x}, \href
  {http://adsabs.harvard.edu/abs/2007MNRAS.381.1197H} {381, 1197}

\bibitem[\protect\citeauthoryear{{Hopkins}, {Bahcall}  \& {Bode}}{{Hopkins}
  et~al.}{2005}]{Hopkins05}
{Hopkins} P.~F.,  {Bahcall} N.~A.,   {Bode} P.,  2005, \mn@doi [\apj]
  {10.1086/425993}, \href {http://adsabs.harvard.edu/abs/2005ApJ...618....1H}
  {618, 1}

\bibitem[\protect\citeauthoryear{{Huang}, {Mandelbaum}, {Freeman}, {Chen},
  {Rozo}, {Rykoff}  \& {Baxter}}{{Huang} et~al.}{2016}]{Huang16}
{Huang} H.-J.,  {Mandelbaum} R.,  {Freeman} P.~E.,  {Chen} Y.-C.,  {Rozo} E.,
  {Rykoff} E.,   {Baxter} E.~J.,  2016, \mn@doi [\mnras]
  {10.1093/mnras/stw1982}, \href
  {http://adsabs.harvard.edu/abs/2016MNRAS.463..222H} {463, 222}

\bibitem[\protect\citeauthoryear{{Joachimi}, {Mandelbaum}, {Abdalla}  \&
  {Bridle}}{{Joachimi} et~al.}{2011}]{Joachimi11}
{Joachimi} B.,  {Mandelbaum} R.,  {Abdalla} F.~B.,   {Bridle} S.~L.,  2011,
  \mn@doi [\aap] {10.1051/0004-6361/201015621}, \href
  {http://adsabs.harvard.edu/abs/2011A%26A...527A..26J} {527, A26}

\bibitem[\protect\citeauthoryear{{Joachimi} et~al.,}{{Joachimi}
  et~al.}{2015a}]{Joachimi15}
{Joachimi} B.,  et~al., 2015a, \mn@doi [\ssr] {10.1007/s11214-015-0177-4},
  \href {http://adsabs.harvard.edu/abs/2015SSRv..193....1J} {193, 1}

\bibitem[\protect\citeauthoryear{{Joachimi}, {Singh}  \&
  {Mandelbaum}}{{Joachimi} et~al.}{2015b}]{Joachimi15b}
{Joachimi} B.,  {Singh} S.,   {Mandelbaum} R.,  2015b, \mn@doi [\mnras]
  {10.1093/mnras/stv1962}, \href
  {http://adsabs.harvard.edu/abs/2015MNRAS.454..478J} {454, 478}

\bibitem[\protect\citeauthoryear{{Kang}, {van den Bosch}, {Yang}, {Mao}, {Mo},
  {Li}  \& {Jing}}{{Kang} et~al.}{2007}]{Kang07}
{Kang} X.,  {van den Bosch} F.~C.,  {Yang} X.,  {Mao} S.,  {Mo} H.~J.,  {Li}
  C.,   {Jing} Y.~P.,  2007, \mn@doi [\mnras]
  {10.1111/j.1365-2966.2007.11902.x}, \href
  {http://adsabs.harvard.edu/abs/2007MNRAS.378.1531K} {378, 1531}

\bibitem[\protect\citeauthoryear{{Kaufmann}}{{Kaufmann}}{1967}]{Kaufmann67}
{Kaufmann} G.~M.,  1967, Some Bayesian Moment Formulae, Report No. 6710. Centre
  for Operations Research and Econometrics, Catholic University of Louvain,
  Heverlee

\bibitem[\protect\citeauthoryear{{Khandai}, {Di Matteo}, {Croft}, {Wilkins},
  {Feng}, {Tucker}, {DeGraf}  \& {Liu}}{{Khandai} et~al.}{2015}]{Khandai15}
{Khandai} N.,  {Di Matteo} T.,  {Croft} R.,  {Wilkins} S.,  {Feng} Y.,
  {Tucker} E.,  {DeGraf} C.,   {Liu} M.-S.,  2015, \mn@doi [\mnras]
  {10.1093/mnras/stv627}, \href
  {http://adsabs.harvard.edu/abs/2015MNRAS.450.1349K} {450, 1349}

\bibitem[\protect\citeauthoryear{{Kiessling} et~al.,}{{Kiessling}
  et~al.}{2015}]{Kiessling15}
{Kiessling} A.,  et~al., 2015, \mn@doi [\ssr] {10.1007/s11214-015-0203-6},
  \href {http://adsabs.harvard.edu/abs/2015SSRv..193...67K} {193, 67}

\bibitem[\protect\citeauthoryear{{Kirk}, {Bridle}  \& {Schneider}}{{Kirk}
  et~al.}{2010}]{Kirk10}
{Kirk} D.,  {Bridle} S.,   {Schneider} M.,  2010, \mn@doi [\mnras]
  {10.1111/j.1365-2966.2010.17213.x}, \href
  {http://adsabs.harvard.edu/abs/2010MNRAS.408.1502K} {408, 1502}

\bibitem[\protect\citeauthoryear{{Kirk}, {Rassat}, {Host}  \& {Bridle}}{{Kirk}
  et~al.}{2012}]{Kirk12}
{Kirk} D.,  {Rassat} A.,  {Host} O.,   {Bridle} S.,  2012, \mn@doi [\mnras]
  {10.1111/j.1365-2966.2012.21099.x}, \href
  {http://adsabs.harvard.edu/abs/2012MNRAS.424.1647K} {424, 1647}

\bibitem[\protect\citeauthoryear{{Kirk} et~al.,}{{Kirk} et~al.}{2015}]{Kirk15}
{Kirk} D.,  et~al., 2015, \mn@doi [\ssr] {10.1007/s11214-015-0213-4}, \href
  {http://adsabs.harvard.edu/abs/2015SSRv..193..139K} {193, 139}

\bibitem[\protect\citeauthoryear{{Koester} et~al.,}{{Koester}
  et~al.}{2007}]{Koester07}
{Koester} B.~P.,  et~al., 2007, \mn@doi [\apj] {10.1086/509599}, \href
  {http://adsabs.harvard.edu/abs/2007ApJ...660..239K} {660, 239}

\bibitem[\protect\citeauthoryear{{Krause}, {Eifler}  \& {Blazek}}{{Krause}
  et~al.}{2016}]{Krause16}
{Krause} E.,  {Eifler} T.,   {Blazek} J.,  2016, \mn@doi [\mnras]
  {10.1093/mnras/stv2615}, \href
  {http://adsabs.harvard.edu/abs/2016MNRAS.456..207K} {456, 207}

\bibitem[\protect\citeauthoryear{{LSST Science Collaboration} et~al.,}{{LSST
  Science Collaboration} et~al.}{2009}]{LSST09}
{LSST Science Collaboration} et~al., 2009, preprint, \href
  {http://adsabs.harvard.edu/abs/2009arXiv0912.0201L} {} (\mn@eprint {arXiv}
  {0912.0201})

\bibitem[\protect\citeauthoryear{{Landy} \& {Szalay}}{{Landy} \&
  {Szalay}}{1993}]{Landy93}
{Landy} S.~D.,  {Szalay} A.~S.,  1993, \mn@doi [\apj] {10.1086/172900}, \href
  {http://adsabs.harvard.edu/abs/1993ApJ...412...64L} {412, 64}

\bibitem[\protect\citeauthoryear{{Laureijs} et~al.,}{{Laureijs}
  et~al.}{2011}]{Laureijs11}
{Laureijs} R.,  et~al., 2011, preprint, \href
  {http://adsabs.harvard.edu/abs/2011arXiv1110.3193L} {} (\mn@eprint {arXiv}
  {1110.3193})

\bibitem[\protect\citeauthoryear{{Li}, {Jing}, {Faltenbacher}  \& {Wang}}{{Li}
  et~al.}{2013}]{Li13}
{Li} C.,  {Jing} Y.~P.,  {Faltenbacher} A.,   {Wang} J.,  2013, \mn@doi [\apjl]
  {10.1088/2041-8205/770/1/L12}, \href
  {http://adsabs.harvard.edu/abs/2013ApJ...770L..12L} {770, L12}

\bibitem[\protect\citeauthoryear{{Mandelbaum}, {Hirata}, {Ishak}, {Seljak}  \&
  {Brinkmann}}{{Mandelbaum} et~al.}{2006}]{Mandelbaum06}
{Mandelbaum} R.,  {Hirata} C.~M.,  {Ishak} M.,  {Seljak} U.,   {Brinkmann} J.,
  2006, \mn@doi [\mnras] {10.1111/j.1365-2966.2005.09946.x}, \href
  {http://adsabs.harvard.edu/abs/2006MNRAS.367..611M} {367, 611}

\bibitem[\protect\citeauthoryear{{Mandelbaum} et~al.,}{{Mandelbaum}
  et~al.}{2011}]{Mandelbaum11}
{Mandelbaum} R.,  et~al., 2011, \mn@doi [\mnras]
  {10.1111/j.1365-2966.2010.17485.x}, \href
  {http://adsabs.harvard.edu/abs/2011MNRAS.410..844M} {410, 844}

\bibitem[\protect\citeauthoryear{{Okumura}, {Jing}  \& {Li}}{{Okumura}
  et~al.}{2009}]{Okumura09}
{Okumura} T.,  {Jing} Y.~P.,   {Li} C.,  2009, \mn@doi [\apj]
  {10.1088/0004-637X/694/1/214}, \href
  {http://adsabs.harvard.edu/abs/2009ApJ...694..214O} {694, 214}

\bibitem[\protect\citeauthoryear{{Paz}, {Sgr{\'o}}, {Merch{\'a}n}  \&
  {Padilla}}{{Paz} et~al.}{2011}]{Paz11}
{Paz} D.~J.,  {Sgr{\'o}} M.~A.,  {Merch{\'a}n} M.,   {Padilla} N.,  2011,
  \mn@doi [\mnras] {10.1111/j.1365-2966.2011.18518.x}, \href
  {http://adsabs.harvard.edu/abs/2011MNRAS.414.2029P} {414, 2029}

\bibitem[\protect\citeauthoryear{{Roche} \& {Eales}}{{Roche} \&
  {Eales}}{1999}]{Roche99}
{Roche} N.,  {Eales} S.~A.,  1999, \mn@doi [\mnras]
  {10.1046/j.1365-8711.1999.02652.x}, \href
  {http://adsabs.harvard.edu/abs/1999MNRAS.307..703R} {307, 703}

\bibitem[\protect\citeauthoryear{{Rozo} \& {Rykoff}}{{Rozo} \&
  {Rykoff}}{2014}]{Rozo14}
{Rozo} E.,  {Rykoff} E.~S.,  2014, \mn@doi [\apj] {10.1088/0004-637X/783/2/80},
  \href {http://esoads.eso.org/abs/2014ApJ...783...80R} {783, 80}

\bibitem[\protect\citeauthoryear{{Rozo}, {Rykoff}, {Bartlett}  \&
  {Melin}}{{Rozo} et~al.}{2015a}]{Rozo15a}
{Rozo} E.,  {Rykoff} E.~S.,  {Bartlett} J.~G.,   {Melin} J.-B.,  2015a, \mn@doi
  [\mnras] {10.1093/mnras/stv605}, \href
  {http://esoads.eso.org/abs/2015MNRAS.450..592R} {450, 592}

\bibitem[\protect\citeauthoryear{{Rozo}, {Rykoff}, {Becker}, {Reddick}  \&
  {Wechsler}}{{Rozo} et~al.}{2015b}]{Rozo15b}
{Rozo} E.,  {Rykoff} E.~S.,  {Becker} M.,  {Reddick} R.~M.,   {Wechsler} R.~H.,
   2015b, \mn@doi [\mnras] {10.1093/mnras/stv1560}, \href
  {http://esoads.eso.org/abs/2015MNRAS.453...38R} {453, 38}

\bibitem[\protect\citeauthoryear{{Rykoff} et~al.,}{{Rykoff}
  et~al.}{2014}]{Rykoff14}
{Rykoff} E.~S.,  et~al., 2014, \mn@doi [\apj] {10.1088/0004-637X/785/2/104},
  \href {http://adsabs.harvard.edu/abs/2014ApJ...785..104R} {785, 104}

\bibitem[\protect\citeauthoryear{{Rykoff} et~al.,}{{Rykoff}
  et~al.}{2016}]{Rykoff16}
{Rykoff} E.~S.,  et~al., 2016, \mn@doi [\apjs] {10.3847/0067-0049/224/1/1},
  \href {http://adsabs.harvard.edu/abs/2016ApJS..224....1R} {224, 1}

\bibitem[\protect\citeauthoryear{{Schaye} et~al.,}{{Schaye}
  et~al.}{2015}]{Schaye15}
{Schaye} J.,  et~al., 2015, \mn@doi [\mnras] {10.1093/mnras/stu2058}, \href
  {http://adsabs.harvard.edu/abs/2015MNRAS.446..521S} {446, 521}

\bibitem[\protect\citeauthoryear{{Simet}, {McClintock}, {Mandelbaum}, {Rozo},
  {Rykoff}, {Sheldon}  \& {Wechsler}}{{Simet} et~al.}{2016}]{Simet16}
{Simet} M.,  {McClintock} T.,  {Mandelbaum} R.,  {Rozo} E.,  {Rykoff} E.,
  {Sheldon} E.,   {Wechsler} R.~H.,  2016, preprint, \href
  {http://adsabs.harvard.edu/abs/2016arXiv160306953S} {} (\mn@eprint {arXiv}
  {1603.06953})

\bibitem[\protect\citeauthoryear{{Singh} \& {Mandelbaum}}{{Singh} \&
  {Mandelbaum}}{2016}]{Singh16IA}
{Singh} S.,  {Mandelbaum} R.,  2016, \mn@doi [\mnras] {10.1093/mnras/stw144},
  \href {http://adsabs.harvard.edu/abs/2016MNRAS.457.2301S} {457, 2301}

\bibitem[\protect\citeauthoryear{{Singh}, {Mandelbaum}  \& {More}}{{Singh}
  et~al.}{2015}]{Singh15}
{Singh} S.,  {Mandelbaum} R.,   {More} S.,  2015, \mn@doi [\mnras]
  {10.1093/mnras/stv778}, \href
  {http://adsabs.harvard.edu/abs/2015MNRAS.450.2195S} {450, 2195}

\bibitem[\protect\citeauthoryear{{Singh}, {Mandelbaum}, {Seljak}, {Slosar}  \&
  {Vazquez Gonzalez}}{{Singh} et~al.}{2016}]{Singh16}
{Singh} S.,  {Mandelbaum} R.,  {Seljak} U.,  {Slosar} A.,   {Vazquez Gonzalez}
  J.,  2016, preprint, \href
  {http://adsabs.harvard.edu/abs/2016arXiv161100752S} {} (\mn@eprint {arXiv}
  {1611.00752})

\bibitem[\protect\citeauthoryear{{Smargon}, {Mandelbaum}, {Bahcall}  \&
  {Niederste-Ostholt}}{{Smargon} et~al.}{2012}]{Smargon12}
{Smargon} A.,  {Mandelbaum} R.,  {Bahcall} N.,   {Niederste-Ostholt} M.,  2012,
  \mn@doi [\mnras] {10.1111/j.1365-2966.2012.20923.x}, \href
  {http://adsabs.harvard.edu/abs/2012MNRAS.423..856S} {423, 856}

\bibitem[\protect\citeauthoryear{{Smith} et~al.,}{{Smith}
  et~al.}{2003}]{Smith03}
{Smith} R.~E.,  et~al., 2003, \mn@doi [\mnras]
  {10.1046/j.1365-8711.2003.06503.x}, \href
  {http://adsabs.harvard.edu/abs/2003MNRAS.341.1311S} {341, 1311}

\bibitem[\protect\citeauthoryear{{Spergel} et~al.,}{{Spergel}
  et~al.}{2015}]{Spergel15}
{Spergel} D.,  et~al., 2015, preprint, \href
  {http://adsabs.harvard.edu/abs/2015arXiv150303757S} {} (\mn@eprint {arXiv}
  {1503.03757})

\bibitem[\protect\citeauthoryear{{Tenneti}, {Singh}, {Mandelbaum}, {Matteo},
  {Feng}  \& {Khandai}}{{Tenneti} et~al.}{2015}]{Tenneti15}
{Tenneti} A.,  {Singh} S.,  {Mandelbaum} R.,  {Matteo} T.~D.,  {Feng} Y.,
  {Khandai} N.,  2015, \mn@doi [\mnras] {10.1093/mnras/stv272}, \href
  {http://adsabs.harvard.edu/abs/2015MNRAS.448.3522T} {448, 3522}

\bibitem[\protect\citeauthoryear{{Tenneti}, {Mandelbaum}  \& {Di
  Matteo}}{{Tenneti} et~al.}{2016}]{Tenneti16}
{Tenneti} A.,  {Mandelbaum} R.,   {Di Matteo} T.,  2016, \mn@doi [\mnras]
  {10.1093/mnras/stw1823}, \href
  {http://adsabs.harvard.edu/abs/2016MNRAS.462.2668T} {462, 2668}

\bibitem[\protect\citeauthoryear{{Troxel} \& {Ishak}}{{Troxel} \&
  {Ishak}}{2015}]{Troxel15}
{Troxel} M.~A.,  {Ishak} M.,  2015, \mn@doi [\physrep]
  {10.1016/j.physrep.2014.11.001}, \href
  {http://adsabs.harvard.edu/abs/2015PhR...558....1T} {558, 1}

\bibitem[\protect\citeauthoryear{{Velliscig} et~al.,}{{Velliscig}
  et~al.}{2015}]{Velliscig15}
{Velliscig} M.,  et~al., 2015, \mn@doi [\mnras] {10.1093/mnras/stv2198}, \href
  {http://adsabs.harvard.edu/abs/2015MNRAS.454.3328V} {454, 3328}

\bibitem[\protect\citeauthoryear{{Wang}, {Park}, {Yang}, {Choi}  \&
  {Chen}}{{Wang} et~al.}{2009}]{Wang09}
{Wang} Y.,  {Park} C.,  {Yang} X.,  {Choi} Y.-Y.,   {Chen} X.,  2009, \mn@doi
  [\apj] {10.1088/0004-637X/703/1/951}, \href
  {http://adsabs.harvard.edu/abs/2009ApJ...703..951W} {703, 951}

\bibitem[\protect\citeauthoryear{{Wang}, {Lin}, {Kang}, {Dutton}, {Yu}  \&
  {Macci{\`o}}}{{Wang} et~al.}{2014}]{Wang14}
{Wang} Y.~O.,  {Lin} W.~P.,  {Kang} X.,  {Dutton} A.,  {Yu} Y.,   {Macci{\`o}}
  A.~V.,  2014, \mn@doi [\apj] {10.1088/0004-637X/786/1/8}, \href
  {http://adsabs.harvard.edu/abs/2014ApJ...786....8W} {786, 8}

\bibitem[\protect\citeauthoryear{{Zu}, {Mandelbaum}, {Simet}, {Rozo}  \&
  {Rykoff}}{{Zu} et~al.}{2016}]{Zu16}
{Zu} Y.,  {Mandelbaum} R.,  {Simet} M.,  {Rozo} E.,   {Rykoff} E.~S.,  2016,
  preprint, \href {http://adsabs.harvard.edu/abs/2016arXiv161100366Z} {}
  (\mn@eprint {arXiv} {1611.00366})

\bibitem[\protect\citeauthoryear{{van Uitert}, {Cacciato}, {Hoekstra}  \&
  {Herbonnet}}{{van Uitert} et~al.}{2015}]{VanUitert15}
{van Uitert} E.,  {Cacciato} M.,  {Hoekstra} H.,   {Herbonnet} R.,  2015,
  \mn@doi [\aap] {10.1051/0004-6361/201525834}, \href
  {http://adsabs.harvard.edu/abs/2015A%26A...579A..26V} {579, A26}

\bibitem[\protect\citeauthoryear{{van Uitert} et~al.,}{{van Uitert}
  et~al.}{2016}]{VanUitert16}
{van Uitert} E.,  et~al., 2016, preprint, \href
  {http://esoads.eso.org/abs/2016arXiv161004226V} {} (\mn@eprint {arXiv}
  {1610.04226})

\makeatother
\end{thebibliography}


\clearpage

\end{document}